\documentclass[journal]{IEEEtran}
\newtheorem{Remark}{\it Remark}[section]

\newtheorem{Proposition}{\it Proposition}[section]

\usepackage{graphicx}
\usepackage{multicol}
\usepackage[T1]{fontenc}
\usepackage{setspace}
\usepackage{amsmath}
\UseRawInputEncoding
\usepackage{epstopdf}
\usepackage{fancyhdr}
\usepackage{indentfirst}
\usepackage{enumerate}
\usepackage{amssymb}
\usepackage{stfloats}
\usepackage{pifont}
\usepackage{multirow}
\usepackage{cleveref}
\usepackage[table]{xcolor}
\usepackage{cite}
\usepackage{subfigure}

\usepackage{algorithm}
\usepackage{algorithmicx}
\usepackage{algpseudocode}
\usepackage{mathrsfs}

\newlength\savewidth

\begin{document}
\title{Capacity Characterization for Reconfigurable Intelligent Surfaces Assisted Multiple-Antenna Multicast}

\author{\IEEEauthorblockN{Linsong Du,~\IEEEmembership{Student Member,~IEEE}, Shihai Shao, \IEEEmembership{Member,~IEEE}, Gang Yang, \IEEEmembership{Member,~IEEE}, Jianhui Ma, \IEEEmembership{Member,~IEEE}, Qingpeng Liang, \IEEEmembership{Member,~IEEE}, and Youxi Tang, \IEEEmembership{Member,~IEEE}}
\thanks{This work was supported by the National Key R\&D Program of China under grant 2018YFB1801903, the National Natural Science Foundation of China under grants 62071093,  61901396, 62071094, U19B2014, 61771107, 61701075, 61601064, 61531009, and Sichuan Science and Technology Program 2020YFH0101.\emph{(Corresponding authors: Shi Hai Shao and Gang Yang)}}

\thanks{L. Du, S. Shao, G. Yang, J. Ma, and Y. Tang are with the National Key Laboratory of Science and Technology on Communications, University of Electronic Science and Technology of China, Chengdu, China, 611731. Emails: linsongdu@outlook.com, ssh@uestc.edu.cn, yanggang@uestc.edu.cn, jianhuima@std.uestc.edu.cn, tangyx@uestc.edu.cn.}

\thanks{Qingpeng Liang is with the School of Information Science and Technology, Southwest Jiaotong University, Chengdu 611756, China (e-mail:
qingpengliang@gmail.com).}

}

\maketitle
\begin{abstract}
 The reconfigurable intelligent surface (RIS), which consists of a large number of  passive and low-cost reflecting elements, has been recognized as a revolutionary technology to enhance the performance of future wireless networks.  This paper considers an RIS assisted multicast transmission, where a base station (BS) with multiple-antenna multicasts common message to multiple single-antenna mobile users (MUs) under the assistance of an RIS.  An equivalent channel model for the considered multicast transmission is analyzed, and then an optimization problem for the corresponding channel capacity is formulated to obtain the optimal covariance matrix and  phase shifts.
In order to solve the above non-convex and non-differentiable problem, this paper first exploits the gradient descent method and alternating optimization, to approach the locally optimal solution for any number of MUs.
Then,  this paper considers a special case, which can obtain the global optimal solution, and shows the sufficient and necessary condition for this special case.
Finally, the order growth of the maximal capacity  is obtained when the numbers of the reflecting elements, the BS antennas, and the MUs go to infinity.
 \end{abstract}


\IEEEdisplaynontitleabstractindextext

\section{Introduction}
In future cellular networks, with the increased demands of sending common messages to multiple mobile users (MUs), such as music sharing, video streaming, and pictures downloading \cite{Cisco2019}, the application of multiple antenna multicast transmission becomes more promising for sending common messages to multiple MUs  with the high transmission rate \cite{7915753}. When compared with the unicast transmission, where the base station (BS) only servers one MU during one time slot, multicast transmission can significantly reduce energy consumption and save spectral resources \cite{7915753}.   However, in many multicast scenarios \cite{6542775}, the direct link from the BS to MUs is blocked, due to the existence of buildings, trees, and cars.  In order to overcome this problem, an effective solution is to add a new link to maintain multicast communication.  Moreover, the capacity of the multicast communication only depends on the minimum received signal-to-noise ratio among all the BS-MU links. It follows that if any one of the links suffers from bad channel conditions, the multicast capacity will become very low. Therefore, it is best to obtain an effective way to improve the BS-MU links with bad channel conditions.


Therefore, the reconfigurable intelligent surface (RIS) technology is considered to be a new way in the multicast communication, which provide wireless connectivity in future 6G system \cite{8869705}. The RIS consists of a large number of reconfigurable reflecting elements, each of which can induce the phase shifts of the electromagnetic waves and then reflects them \cite{1903-08925,8466374,9316920,9110869}. The RIS can proactively control the multicast channel between the BS and MUs via highly controllable and intelligent signal reflection. Thus, the RIS provides a new degree of freedom to improve the performance of the multicast communication.
  It is worth noting that the RIS can be regarded as a no-power  full-duplex (FD) amplify-and-forward (AF) relay with multiple-antenna, which receives signals and then forwards them to the MUs. However, there exists some differences between the relay and the RIS as follows.  Similar to backscatter \cite{8103807,8274950,8424210}, since the RIS does not use the transmit radio frequency (RF) chain, it hardly consumes any energy. Thus, RIS can achieve much higher energy efficiency and is more environmentally-friendly than the regular  FD AF relay, and does not cause self-interference as well.
In fact,  the RIS can achieve a higher rate than the AF relay when the total transmit power (i.e., the sum of the transmit power at the BS and the AF relay) is fixed  \cite{9119122}.
Moreover, comparing with massive multiple-input multiple-output (MIMO), the RISs have a less complex structure and lower cost. They are easy to densely deploy at various types of places such as trees, buildings, and rooms \cite{8319526}.   The RIS is able to cater to the different application scenarios. First, the RIS can provide a new link to maintain transmission in the dead zone \cite{8888223,8746155,li2019reconfigurable,8741198}, where the direct paths between the BS and the MUs are blocked by obstacles. Then, the RIS can improve the physical layer security by enhancing the desired signals and suppressing the undesired signals \cite{8723525,8847342,8649589,9112318}. In addition, when MU suffers co-channel interference from other interferers, the RIS can be placed to suppress the interference.
Finally, indoor environments can be coated with the RIS to increase the throughput offered by conventional access points.
Due to the above benefits,  the RIS can adaptively adjust the phase shifts of the received signals, it can enhance the BS-MU links with bad channel conditions under indoor and outdoor scenarios. Therefore, this paper focuses on RIS assisted multicast communication.

However, the application of the RIS on multicast communication is confronted with a challenge. In order to optimize the phase shifts at the RIS, the RIS assisted multicast system requires the accurate channel state information (CSI) on the RIS-related channel with the BS and the MUs. The acquisition of CSI on the RIS-related channel is difficult since RIS without any RF chains cannot perform baseband processing functionality. Therefore, the CSI of BS-RIS and RIS-MUs links cannot be separately estimated via the traditional training-based approaches in general \cite{9180053,9366805}. Considering this issue, the authors in \cite{zhang2019capacity} considered deploying the dedicated RF chains at RIS to acquire the CSI of RIS-MUs links. However, this approach increases the implementation cost and decreases energy efficiency, which loses the essential benefits of the RIS.
Thus, the authors in \cite{9154276,9130088,8937491,9261597,9195133} proposed the channel estimation methods in multiple MUs scenario to obtain the cascaded CSI of BS-RIS-MU links without using RF chains at the RIS. In \cite{9154276,9130088}, the BS can obtain the global CSI of all BS-RIS-MU links by uplink channel estimation due to channel reciprocity, where each MU transmits its pilot symbols to the BS on the different time slots, such that each cascaded CSI of BS-RIS-MU link is estimated by BS. A three-phase channel estimation framework was proposed in \cite{9130088} to shorten the estimated time.
In \cite{8937491}, a transmission protocol was proposed to estimate cascaded CSI of BS-RIS-MU for orthogonal frequency division multiplexing (OFDM) system under unit-modulus constraint.  In \cite{9261597}, a fast channel estimation scheme with reduced OFDM symbol duration was proposed for the fading channel. In \cite{9195133}, the SeUCE scheme was proposed to estimate cascaded CSI for multi-user OFDM access system. The cascaded CSI of the BS-RIS-MU link is sufficient for the design of the phase shifts at the RIS and the covariance matrix at the BS.

Based on the available cascaded CSI, the authors \cite{Jun2019,9076830,9110869} studied how to optimize the phase shifts at the RIS in a multicast system.  The authors in \cite{Jun2019} considered the fair quality of service for multicasting assisted by the RIS and proposed efficient algorithms to optimize the quality of service by jointly designing the transmit beamforming and the phase shifts.
The authors in \cite{9076830} considered the RIS assisted multi-group multicasting and maximized the sum rate of all the multicasting groups by optimizing the transmit beamforming and the phase shifts.
 However, the works in \cite{Jun2019,9076830}  only study the transmit beamforming vector at BS, which cannot embody the theoretically maximal achievable rate, i.e., capacity, for RIS assisted multicast communication. The study of capacity indicates the optimal performance achievable on the RIS assisted multicast channel and how to achieve such optimal performance. In order to obtain the capacity, the covariance matrix at BS should be optimized, which brings a new challenge. Indeed, the authors in \cite{zhang2019capacity} studied the capacity characterization for the RIS assisted MIMO communications in the unicast scenario.  However, the results in \cite{zhang2019capacity} cannot generalize to the multicast scenario. The capacity maximization problem in the multicast scenario is more difficult to solve as compared to that in the unicast scenario since the capacity maximization problem in RIS assisted multicast communication is a non-differentiable max-min problem, and phase shifts and covariance matrix need to be designed to balance the different BS-RIS-MU links. To the best of our knowledge, there is no work considering capacity characterization for the RIS assisted multicast communication.

In this paper, we consider the RIS assisted multi-antenna multicast transmission, where a multi-antenna BS  sends common messages to a group of single-antenna MUs.  An RIS consisting of a large number of reconfigurable reflecting elements is deployed to assist the multicast transmission, where the BS sends signals to the RIS, and then the RIS  forwards the received signals to the MUs with phase shifts. It is assumed that the cascaded CSI of BS-RIS-MU links and the CSI of BS-MUs links are perfectly known to the BS and the RIS. The equivalent channel model for the considered multicast system is obtained, which can be regarded as a conventional multicast channel, and its characteristics can be partly controlled by the reconfigurable reflecting elements.  Thus, it is crucial to find the optimal phase shifts of the RIS to maximize the capacity of the equivalent channel model. The main contributions of this paper are summarized as follows.
\begin{itemize}
\item  First, an optimization problem of the channel capacity is formulated to obtain the optimal phase shifts for the RIS and the corresponding covariance matrix of the transmitted symbol vector for the BS. Since this problem is non-convex and non-differentiable, it is difficult to obtain the optimal solution directly. Thus, the non-differentiable problem is reformulated into a differentiable problem. Then, the gradient descent method and alternating optimization, respectively, are proposed to approach a locally optimal solution. We consider a specific case, which owns optimal semi-closed form solution, and show the sufficient and necessary conditions that the special case happens.
\item Next,  we analyze the order growth of the optimal capacity of  RIS assisted multicast transmission in some asymptotic cases: 1) The number of MUs is fixed and the number of antennas or reflecting elements goes to infinity; 2) The numbers of antennas and reflecting elements are fixed, and the number of MUs goes to infinity; 3) The numbers of antennas, reflecting elements, and MUs all go to infinity.
\item Finally, we numerically evaluate the performance of the proposed two algorithms, which confirm the asymptotic analysis of optimal capacity. From the numerical results and asymptotic analysis, we observe that the optimal capacity grows logarithmically with the number of antennas and the square of the number of reflecting elements, and it is the inverse proportion with the number of MUs. When both numbers of MUs and antennas go to infinity at a fixed ratio, the optimal capacity remains a constant.
\end{itemize}

The structure of this paper is organized as follows. Section \ref{section_system} introduces the system model and formulates an optimization problem of the channel capacity. Section \ref{section_km} proposes two algorithms for optimization problem for any number of MUs.  Section \ref{section_asymptotic} shows the asymptotic analysis for the capacity of the RIS assisted multicast transmission.  Section \ref{section_Numerical} presents some numerical results. Section \ref{section_conclusion} concludes the paper.

\emph{Notation}: $\mathbf{a}$ is a vector,  $\mathbf{A}$ is a matrix.
$\left\|\mathbf{a} \right\|$ is Euclidean norm of  $\mathbf{a}$.
$\mathbf{A}^H$, $\mathbf{A}^T$, $\left\|\mathbf{A} \right\|_{m_2}$, $\mathbf{A}^{-1}$  denote Hermitian transpose, transpose, Frobenius norm and pseudo-inverse of $\mathbf{A}$, respectively.
$\mathrm{diag}\left(\mathbf{a}\right)$ is a diagonal matrix with the entries of $\mathbf{a}$ on its main diagonal.
$\mathbf{A} \succeq \mathbf{B}$ means that $\mathbf{A}-\mathbf{B}$ is positive semidefinite.
$\mathbf{A}\otimes\mathbf{B}$ denotes the kronecker product between $\mathbf{A}$ and $\mathbf{B}$.
 $\mathrm{Tr}\left(\mathbf{A}\right)$ denotes the trace of $\mathbf{A}$.
$\mathbb{E}\left\{\mathbf{A}\right\}$ denotes the expected value of  each element for $\mathbf{A}$.
$\det\mathbf{A}$ denotes the determinant of $\mathbf{A}$.
$\text{vec}\left(\mathbf{A}\right)$ is an operator that transforms $\mathbf{A}$ into a column vector by vertically stacking the columns of the matrix.
$\frac{\mathbf{A}}{x}$ means that the each element of  $\mathbf{A}$ divides by $x$.
 $\mathbb{C}^{N \times M}$ denotes the set of all $N\times M$ complex-valued matrices.
$\mathbf{I}$ is an identity matrix.
$j \triangleq \sqrt{-1}$   is the imaginary unit.
$\arg\left(\cdot\right)$ denotes the argument of a complex number.
$\mathbf{Cov}\left(\cdot,\cdot\right)$ is covariance.
\section{System Model and Problem Formulation} \label{section_system}

\subsection{System Model}
\begin{figure}[!t]
\centering\includegraphics[width=3in]{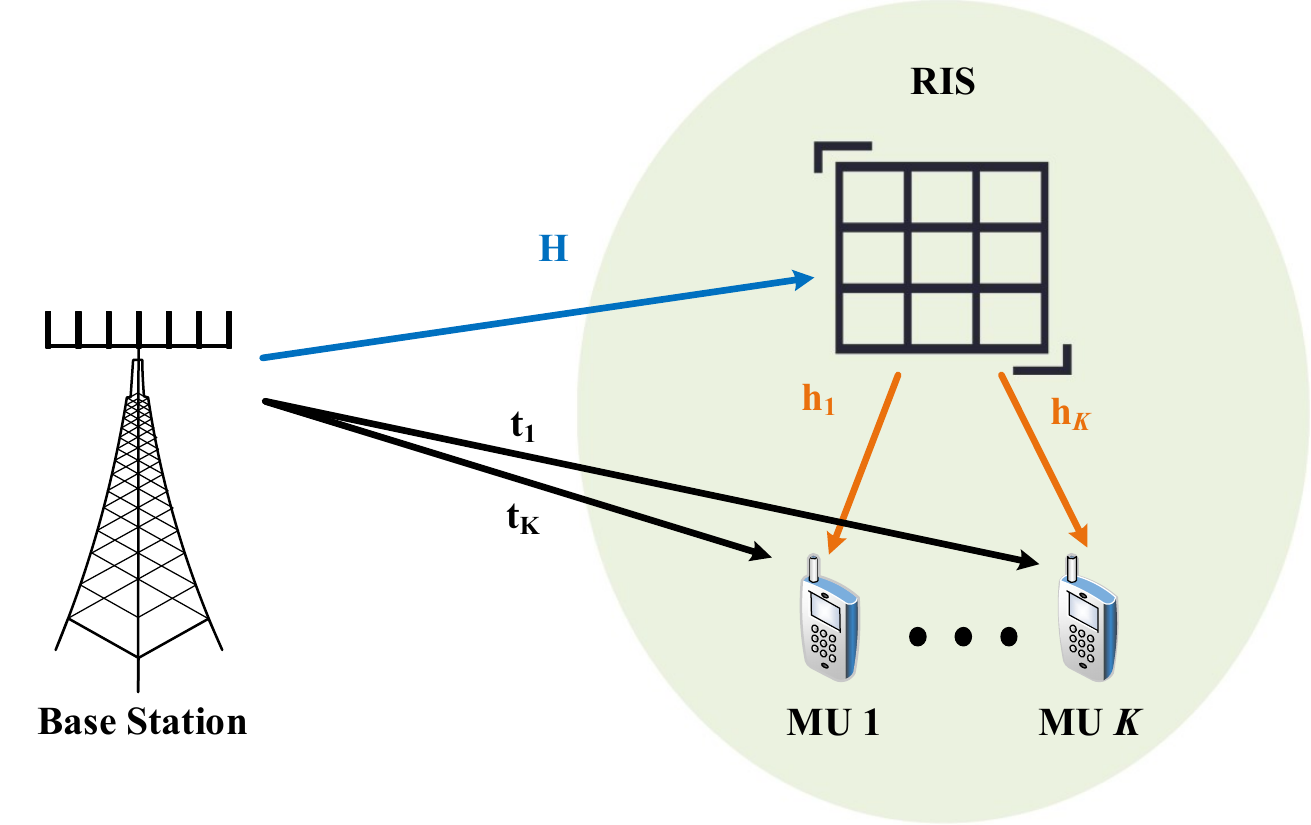}
\caption{RIS assisted multicast transmissions with $K$ MUs.}\label{fig_system}
\end{figure}
As shown in Fig. \ref{fig_system}, a BS with $M$-antennas sends a common message to $K$ single-antenna MUs\footnote{Since we focus on the capacity for RIS assisted multicast, we consider the single group scenario (pure multicast), i.e.,  the BS sends the same message to all $K$ MUs. If this is the multi-group scenario ($K$ MUs divided into multiple groups, and the BS sends an independent message to each group), we need to study the capacity region rather than capacity.} \cite{4036286,4524056}.  An RIS with $N$ reflecting elements is deployed between the BS and the $K$ MUs, where the cascaded CSI of BS-RIS-MU link and the CSI of BS-MU link are perfectly known.   Accordingly, the received signal at MU $k$, $k=1,2,\cdots,K$ is given by
\begin{align}
y_{k}=\left(\mathbf{h}_{k}^{H}\mathbf{\Phi} \mathbf{H}+\mathbf{t}_{k}^{H}\right)\mathbf{s}+z_k, \label{received_signal_1}
\end{align}
where  $\mathbf{s}=\left[s_1,s_2,\cdots,s_M\right]^T$  is the transmitted symbol vector;
$\mathbf{t}_{k}^{H}\in\mathbb{C}^{1 \times M}$  is the channel vector between the BS and MU $k$, i.e.,
 $\mathbf{t}_{k}^{H}=\left[t_{k,1},t_{k,2},\cdots,t_{k,M}\right]$,
with $t_{t,m}=a_{t_{t,m}}e^{j\theta_{t_{k,m}}}$;
$\mathbf{H}\in \mathbb{C}^{N \times M}$ is the channel matrix between the BS and the RIS, i.e.,
\begin{align}
\mathbf{H}=\begin{bmatrix}
H_{1,1} & H_{1,2}  & \cdots  &  H_{1,M}  \\
H_{1,2} & H_{2,2}& \cdots & H_{2,M}\\
\vdots  &    \vdots &  \ddots &\vdots \\
 H_{1,N}&  H_{2,N} &\cdots  & H_{N,M}\\
\end{bmatrix},\label{H}
\end{align}
 with $H_{n,m}=a_{H_{n,m}}e^{j\theta_{H_{n,m}}}$;
$\mathbf{h}_{k}^{H}\in\mathbb{C}^{1 \times N}$ is the channel vector between the RIS and the MU $k$, i.e.,
\begin{align}
 \mathbf{h}_{k}^{H}=\left[h_{k,1},h_{k,2},\cdots,h_{k,N}\right],\label{h}
\end{align}
with $h_{k,n}=a_{h_{k,n}}e^{j\theta_{h_{k,n}}}$;
$\mathbf{\Phi}=\mathrm{diag}\left[\Phi_1, \Phi_2,\cdots, \Phi_N \right]\in \mathbb{C}^{N \times N}$  represents the phase shifts introduced by the reflecting elements at the RIS, with $\Phi_n=e^{j\theta_n}$;
and $z_k$ is the circularly symmetric complex Gaussian (CSCG) noise with zero mean and unit variance $\sigma^2=1$.

In fact, it is hard to directly obtain the CSI of the separate BS-RIS  and separate RIS-MU links, i.e., $\mathbf{H}$ and $\mathbf{h}_{k}^{H}$.  However,  the signal model (\ref{received_signal_1}) is equivalent to
\begin{align}
y_{k}=\left(\mathbf{u}^{H}\mathbf{G}_k \mathbf{H}+\mathbf{t}_{k}^{H}\right)\mathbf{s}+z_k, \label{received_signal_2}
\end{align}
where $\mathbf{G}_k \mathbf{H}$ is the cascaded channel form the BS to the $k$-th MU via RIS, and $\mathbf{u}^{H}=\left[\Phi_1, \cdots, \Phi_N\right]$, $\mathbf{G}_k=\mathrm{diag}\left(\mathbf{h}_k\right)$. Notice that the cascaded CSI of the channel matrix $\mathbf{G}_k \mathbf{H}$ can be acquired to the BS, which implies that the each element of matrix $\mathbf{G}_k \mathbf{H}$, i.e., $h_{k,n}H_{n,m}=a_{h_{k,n}}a_{H_{n,m}}e^{j(\theta_{h_{k,n}}+\theta_{H_{n,m}})}$, is known.

\subsection{Problem Formulation}
For the fixed $\mathbf{\Phi}$,  the capacity of RIS assisted multicast channel is given by \cite{4036286,4524056}
\begin{align}
C\left(\mathbf{\Phi}\right)\triangleq \max_{\mathbf{Q}:\mathbf{Q}\succeq 0, \mathrm{Tr}\left(\mathbf{Q}\right)\leq P_{\max}} & \min_{k=1,\cdots, K}\log \Big[1+\nonumber \\
&\left.\left(\mathbf{u}^{H}\mathbf{G}_k \mathbf{H}+\mathbf{t}_{k}^{H}\right)\mathbf{Q}\left(\mathbf{H}^{H}\mathbf{G}^{H}_k \mathbf{u}+\mathbf{t}_{k}\right)\right]
\label{Capacity_formulaiton}
\end{align}
where  $\mathbf{Q}=\mathbb{E}\left\{\mathbf{s}\mathbf{s}^H\right\}$ is the covariance matrix of the transmitted symbol vector  $\mathbf{s}$;   $P_{\max}$ is the power budget, and  $\mathrm{Tr}\left(\mathbf{Q}\right)\leq P_{\max}$ is the power constraint.
From (\ref{Capacity_formulaiton}),  it is observed that the capacity depends on $\mathbf{\Phi}$. In order to obtain the optimal capacity for the equivalent multicast channel by jointly the optimizing the covariance matrix $\mathbf{Q}$ and the phase shifts $\boldsymbol{\theta}=\left[\theta_1,\theta_2,\cdots,\theta_N\right]^T$, an optimization problem is formulated as
\begin{align}
(\textrm{P1})~~C =\max_{\mathbf{Q},\boldsymbol{\theta}}& \min_{k=1,\cdots, K}\log\Big[1+\nonumber\\
&~~~~~~~~~~~~~\left.\left(\mathbf{u}^{H}\mathbf{G}_k \mathbf{H}+\mathbf{t}_{k}^{H}\right)\mathbf{Q}\left(\mathbf{H}^{H}\mathbf{G}^{H}_k \mathbf{u}+\mathbf{t}_{k}\right)\right]\label{Problem_1}\\
\textrm{s.t.} ~& \mathrm{Tr}\left(\mathbf{Q}\right)\leq P_{\max},\label{SJ_QP}\\
&\mathbf{Q}\succeq 0,  \label{SJ_Q}\\
&\left|\Phi_n\right|=1, n=1,2,\cdots,N.\label{SJ_theta}
 \end{align}

Since the objective function in (\ref{Problem_1}) is non-convex due to the phase shifts \cite{IEEEhowto:S.Boyd}, Problem (P1) is non-convex, and its optimal solution cannot be obtained directly by any effective and standard method. Moreover, the objective function (\ref{Problem_1}) is non-differentiable because the pointwise minimum $\min\left\{\cdot\right\}$  is non-differentiable \cite{IEEEhowto:S.Boyd}, and hence the KKT conditions\footnote{KKT conditions are the necessary conditions that optimal solutions need to satisfy for a differentiable problem.} for (P1) do not exist \cite{IEEEhowto:S.Boyd}.


\section{Proposed Algorithm to Problem (P1)}\label{section_km}
 In this section, we propose two efficient algorithms to find the phase shifts and covariance matrix for Problem (P1).
First, the objective function in (\ref{Problem_1}) for Problem (P1) is non-differentiable,  and thus we reformulate  (P1) as the following differentiable problem:
\begin{align}
(\textrm{P2})~~\max_{\mathbf{Q},\boldsymbol{\theta},\gamma}&~~\gamma\\
\textrm{s.t.} ~&\gamma\leq\left(\mathbf{u}^{H}\mathbf{G}_k \mathbf{H}+\mathbf{t}_{k}^{H}\right)\mathbf{Q}\left(\mathbf{H}^{H}\mathbf{G}^{H}_k \mathbf{u}+\mathbf{t}_{k}\right), \nonumber\\
&k=1,\cdots,K, \label{constraint_Problem3}\\
& (\ref{SJ_QP}),(\ref{SJ_Q}), (\ref{SJ_theta}), \nonumber
 \end{align}
where $\gamma$ is an auxiliary variable. Notice that Problems  (P1) and  (P2) have the same optimal solution $\left\{\mathbf{Q}_*,\boldsymbol{\theta}_*\right\}$.  In this section, we propose two algorithms, i.e., gradient descent method and alternating optimization to obtain the locally optimal solution for  (P2).

Since (P2) is an inequality constrained optimization problem and continuously differentiable, we perform the logarithmic barrier method, which is one of the descent methods \cite{IEEEhowto:S.Boyd}, to converge the objective function of (P2) as a local optimum.

\subsection{Gradient Descent Method}
We reformulate Problem (P2) as an unconstrained minimization problem
\begin{align}
(\text{P2-t})~&\min \Gamma^{\left(t\right)}\left(\mathbf{Q}, \boldsymbol{\theta},\gamma\right), \label{Problem_3*}
 \end{align}
where $ \Gamma^{\left(t\right)}\left(\mathbf{Q}, \boldsymbol{\theta},\gamma\right)=-\gamma-\frac{1}{t}f\left(\mathbf{Q}, \boldsymbol{\theta},\gamma\right)$, and $f\left(\mathbf{Q}, \boldsymbol{\theta},\gamma\right)$ is given as
\begin{align}
f\left(\mathbf{Q}, \boldsymbol{\theta},\gamma\right)=&\sum_{k=1}^{K}\log\left(-\gamma+\left(\mathbf{u}^{H}\mathbf{G}_k \mathbf{H}+\mathbf{t}_{k}^{H}\right)\mathbf{Q}\left(\mathbf{H}^{H}\mathbf{G}^{H}_k \mathbf{u}+\mathbf{t}_{k}\right)\right)\nonumber\\
&+\log\left(-\mathrm{Tr}\left(\mathbf{Q}\right)+P_{\max}\right)+\log\left(\det\mathbf{Q}\right),
 \end{align}
and  $-\frac{1}{t}\log(-x)$ is the logarithmic barrier function for the inequality constraints.

Problem  (P2-t)   can be regarded as an approximation of Problem (P2), where $t>0$ is a parameter to the accuracy of the approximation.  Thus, a large value of $t$ can be used to approximate Problem (P2).

We solve a sequence of problems with each corresponding to (P2-t)  for a certain value of  $t$ sorted in ascending order, and the optimal point for the previous problem in the sequence  is used as the initial value for the current one \cite{IEEEhowto:S.Boyd}.

For Problem  (P2-t), we perform the gradient descent method to compute the optimal solution  $\left\{\mathbf{Q}^{\left(t\right)},\boldsymbol{\theta}^{\left(t\right)},\gamma^{\left(t\right)}\right\}$, where the descent direction $\left\{\Delta\mathbf{Q},\Delta\boldsymbol{\theta},\Delta\gamma\right\}$ and the step size $k$  are obtained as follows.
\begin{itemize}
	\item \textbf{Descent Direction:}   By taking the derivative of $\Gamma^{\left(t\right)}\left(\mathbf{Q}, \boldsymbol{\theta},\gamma\right)$ with respect to $\mathbf{Q}$, $\boldsymbol{\theta}$, and $\gamma$, respectively, we obtain the descent direction as
\begin{align}
\Delta\mathbf{Q}=&\frac{1}{t}\sum_{k=1}^{K}\frac{\left(\mathbf{H}^{H}\mathbf{G}^{H}_k \mathbf{u}+\mathbf{t}_{k}\right)\left(\mathbf{u}^{H}\mathbf{G}_k \mathbf{H}+\mathbf{t}_{k}^{H}\right)}{-\gamma+\left(\mathbf{u}^{H}\mathbf{G}_k \mathbf{H}+\mathbf{t}_{k}^{H}\right)\mathbf{Q}\left(\mathbf{H}^{H}\mathbf{G}^{H}_k \mathbf{u}+\mathbf{t}_{k}\right)}\nonumber\\
&+\frac{\mathbf{I}}{t\left(\mathrm{Tr}\left(\mathbf{Q}\right)-P_{\max}\right)}+\frac{\mathbf{Q}^{-1}}{t},\label{direction_Q_K3}\\
\Delta\boldsymbol{\theta}=&\frac{1}{t}\sum_{k=1}^{K}\frac{\left(\mathbf{G}_{k}\mathbf{H}\mathbf{Q}\mathbf{H}^{H}\mathbf{G}_k ^{H}+2\mathbf{t}_{k}^{H}\mathbf{Q}\mathbf{H}^{H}\mathbf{G}_k ^{H}\right)\boldsymbol{\varphi}}{-\gamma+\left(\mathbf{u}^{H}\mathbf{G}_k \mathbf{H}+\mathbf{t}_{k}^{H}\right)\mathbf{Q}\left(\mathbf{H}^{H}\mathbf{G}^{H}_k \mathbf{u}+\mathbf{t}_{k}\right)},\label{direction_theta_K3}\\
\Delta \gamma=&1-\frac{1}{t}\sum_{k=1}^{K}\frac{1}{-\gamma+\left(\mathbf{u}^{H}\mathbf{G}_k \mathbf{H}+\mathbf{t}_{k}^{H}\right)\mathbf{Q}\left(\mathbf{H}^{H}\mathbf{G}^{H}_k \mathbf{u}+\mathbf{t}_{k}\right)}.\label{direction_T_K3}
\end{align}
where $\boldsymbol{\varphi}=\left[je^{j\theta_{1}},\cdots,je^{\theta_{N}}\right]^T$.

\item  \textbf{Step Size:} We also adopt the backtracking line search to determine the step size $k$, where the initial value of $k$ is $k=1$ and then the value of  $k$ is reduced by $k:= k\eta $, with $0<\eta<1$, until the stopping condition
\begin{align}
& \Gamma^{\left(t\right)}\left(\mathbf{Q}+k\Delta\mathbf{Q}, \boldsymbol{\theta}+k\Delta\boldsymbol{\theta},\gamma+k\Delta \gamma\right)\nonumber\\
<&\Gamma^{\left(t\right)}\left(\mathbf{Q}, \boldsymbol{\theta}, \gamma\right)- \alpha k\left(\left\|\Delta\mathbf{Q} \right\|_{m_2}^2+\left\|\Delta\boldsymbol{\theta}\right\|^2+ \left|\Delta \gamma\right|^2\right)\label{condition_step_direct}
\end{align}
satisfies.
\end{itemize}

\textbf{Algorithm Summary:} Based on the above discussion, we can compute the optimal solution $\left\{\mathbf{Q}_*,\boldsymbol{\theta}_*,\gamma_*\right\}$ for Problem (P2) by two-level iterations, and the detailed algorithm is summarized in Algorithm \ref{table_P2}.
\begin{algorithm}
        \caption{ Gradient Descent  Method for Problem (P2) }
\label{table_P2}
        \begin{algorithmic}[1]
\Require  $\mathbf{G}_k$, $k=1,\cdots, K$, $\mathbf{H}$, $P_{\max}$, and the error tolerances $\delta_1 >0$ and $\delta_2>0$.
 \Ensure $\left\{\mathbf{Q}_{*},\boldsymbol{\theta}_{*}, \gamma_{*}\right\}$.
             \State {\bf Initialize} $\left\{\mathbf{Q}^{\left(t_i\right)},\boldsymbol{\theta}^{\left(t_i\right)},\gamma^{\left(t_i\right)}\right\}$ and $\left\{\mathbf{Q}_{j},\boldsymbol{\theta}_{j},\gamma_{j}\right\}$, which represent the output of the $i$-th  outer iteration and the input of the $j$-th  inner iteration, respectively.
 \While{$\frac{1}{t_{i-1}}>\delta_1 $}
\State Let $t_{i}=\rho t_{i-1}$.
\State {\bf Initialize} $\mathbf{Q}=\mathbf{Q}^{\left(t_{i-1}\right)}$, $\boldsymbol{\theta}=\boldsymbol{\theta}^{\left(t_{i-1}\right)}$, and $\gamma^{\left(t_{i-1}\right)}$.
\While{$\left|\Gamma^{\left(t_{i}\right)}\left(\mathbf{Q}_{j}, \boldsymbol{\theta}_{j},\gamma_{j}\right)-\Gamma^{\left(t_{i}\right)}\left(\mathbf{Q}_{j-1}, \boldsymbol{\theta}_{j-1},\gamma_{j-1}\right)\right|>\delta_2$}
\State  Compute $\Delta\mathbf{Q}_{j+1}$, $\Delta\boldsymbol{\theta}_{j+1}$, and  $\Delta \gamma_{j+1}$ by (\ref{direction_Q_K3}), (\ref{direction_theta_K3}) and (\ref{direction_T_K3}), respectively.
\State {\bf Initialize} $l_{1}=l_{0}$
\While {Condition (\ref{condition_step_direct}) is false}
 \State Let $l_{i}:=l_{i-1}\eta$.
\EndWhile
\State Let $\mathbf{Q}_{j+1}=\mathbf{Q}_{j}+l_{i}\Delta\mathbf{Q}_{j+1}$,  $\boldsymbol{\theta}_{j+1}=\boldsymbol{\theta}_{j}+l_{i}\Delta\boldsymbol{\theta}_{j+1}$, and $\gamma_{j+1}=\gamma_{j}+l_{i}\Delta \gamma_{j+1}$.
\EndWhile
\State Let $\mathbf{Q}^{\left(t_{i}\right)}=\mathbf{Q}_{j+1}$, $\boldsymbol{\theta}^{\left(t_{i}\right)}=\boldsymbol{\theta}_{j+1}$, and $\gamma^{\left(t_{i}\right)}=\gamma_{j+1}$.
\EndWhile
\State Let  $\left\{\mathbf{Q}_*,\boldsymbol{\theta}_*,\gamma_{*}\right\}=\left\{\mathbf{Q}^{\left(t_i\right)},\boldsymbol{\theta}^{\left(t_i\right)},\gamma^{\left(t_i\right)}\right\}$.
  \end{algorithmic}
\end{algorithm}

From  (\ref{condition_step_direct}), it follows that
$\Gamma^{\left(t\right)}\left(\mathbf{Q}, \boldsymbol{\theta}\right)$ decreases over iterations. Thus, Algorithm \ref{table_P2}
 is guaranteed to converge to a locally optimal solution.  The convergence rate of the Algorithm \ref{table_P2} is at  least linear form \cite{IEEEhowto:S.Boyd}.

 From \cite{Nemirovski2008}, we obtain that the complexity of each iteration is mainly due to the calculations of $\Delta\mathbf{Q}$ and $\Delta\boldsymbol{\theta}$, whose complexities are $\mathcal{O}\left(\left(MN^2+M^2N\right)K+M^3\right)$ \cite{Nemirovski2008} and $\mathcal{O}\left(N^2+M^2+MN\right)$.  It is obvious that the order of the complexity of  $\Delta\boldsymbol{\theta}$  is less than that of $\Delta\mathbf{Q}$, we can ignore the complexity of  $\Delta\boldsymbol{\theta}$. Then, the number of the inner and outer iterations are $I$ and $J$, respectively, and thus the complexity of Algorithm \ref{table_P2} is given as  $\mathcal{O}\left(\left(\left(MN^2+M^2N\right)K+M^3\right)IJ\right)$ \cite{Nemirovski2008}.

\subsection{Alternating Optimization}
If we only optimize $\mathbf{Q}$ for Problem (P2), the corresponding subproblem is convex. Thus, we consider adopting the alternating optimization technique to separately and iteratively solve for $\mathbf{Q}$ and $\boldsymbol{\theta}$. In each iteration, we first optimize $\mathbf{Q}$ with fixed $\boldsymbol{\theta}$, and then solve for $\boldsymbol{\theta}$ with fixed $\mathbf{Q}$.
For the phase shift optimization problem,  the non-convex constraints (\ref{constraint_Problem3}) are handled by the characteristic of multicast capacity expression, and then we obtain the optimal semi-closed form solution.

\subsubsection{Optimization of $\mathbf{Q}$ with Fixed $\boldsymbol{\theta}$}
For the fixed $\boldsymbol{\theta}$, the subproblem which only optimizes  $\mathbf{Q}$ is formulated as
\begin{align}
(\textrm{P2.1})~~\max_{\mathbf{Q},\gamma}&~~\gamma\\
\textrm{s.t.} ~&\gamma\leq \mathrm{Tr}\left(\mathbf{R}_{k}\mathbf{Q}\right), k=1,\cdots,K \label{constraint_Problem3-1}\\
& (\ref{SJ_QP}),(\ref{SJ_Q}), \nonumber
 \end{align}
where $\mathbf{R}_{k}=\left(\mathbf{H}^{H}\mathbf{G}^{H}_k \mathbf{u}+\mathbf{t}_{k}\right)\left(\mathbf{u}^{H}\mathbf{G}_k \mathbf{H}+\mathbf{t}_{k}^{H}\right)$, and (\ref{constraint_Problem3-1}) is due to
\begin{align}
 \left(\mathbf{u}^{H}\mathbf{G}_k \mathbf{H}+\mathbf{t}_{k}^{H}\right)\mathbf{Q}\left(\mathbf{H}^{H}\mathbf{G}^{H}_k \mathbf{u}+\mathbf{t}_{k}\right)=\mathrm{Tr}\left(\mathbf{R}_{k}\mathbf{Q}\right).
 \end{align}
Obviously, Problem (P2) is a convex problem and can be efficiently solved by standard semi-definite programming (SDP)  \cite{IEEEhowto:S.Boyd} and we can employ standard convex optimization tools, e.g., CVX \cite{cvx}, to compute the optimal solution.

\subsubsection{Optimization of $\boldsymbol{\theta}$ with Fixed $\mathbf{Q}$}
Next, we solve the following subproblem with the fixed  $\mathbf{Q}$:
\begin{align}
(\textrm{P2.2})~~\max_{\boldsymbol{\theta},\gamma}&~~\gamma\label{Problem_2-2}\\
\textrm{s.t.} ~&(\ref{SJ_theta}), (\ref{constraint_Problem3}). \nonumber
 \end{align}
Since $\mathbf{Q}$ is a positive semi-definite matrix, the eigenvalue decomposition of $\mathbf{U}\Sigma \mathbf{U}^{H}$ can be obtained, where $\mathbf{U}\in\mathbb{C}^{M\times M}$, and diagonal elements in $\Sigma$ are non-negative real numbers.
Thus, we can define $\mathbf{V}=\mathbf{H}\mathbf{U}\Sigma^{\frac{1}{2}}=\left[\mathbf{v}_1,\cdots,\mathbf{v}_N\right]^{H}\in \mathbb{C}^{N\times M}$ and $\mathbf{p}_k=\Sigma^{\frac{1}{2}}\mathbf{U}^{H}\mathbf{t}_k\in\mathbb{C}^{M\times 1}$. Then, $\left(\mathbf{u}^{H}\mathbf{G}_k \mathbf{H}+\mathbf{t}_{k}^{H}\right)\mathbf{Q}\left(\mathbf{H}^{H}\mathbf{G}^{H}_k \mathbf{u}+\mathbf{t}_{k}\right)$ in (\ref{constraint_Problem3}) can be rewritten as
\begin{align}
&\left(\mathbf{u}^{H}\mathbf{G}_k \mathbf{H}+\mathbf{t}_{k}^{H}\right)\mathbf{Q}\left(\mathbf{H}^{H}\mathbf{G}^{H}_k \mathbf{u}+\mathbf{t}_{k}\right)\nonumber\\
=&\mathbf{u}^{H}\mathbf{G}_k\mathbf{V}\mathbf{V}^H\Phi^{H}\mathbf{t}_{k}+\mathbf{u}^{H}\mathbf{G}_k\mathbf{V}\mathbf{p}_k+\mathbf{p}_k^H\mathbf{V}^H\mathbf{G}_k^{H}\mathbf{u}+\mathbf{t}_k^{H}\mathbf{Q}\mathbf{t}_k\nonumber\\
=&\sum_{i=1}^{N}e^{j\theta_i}h_{k,i}\mathbf{v}_i^H\sum_{j=1}^{N} e^{-j\theta_j}h_{k,j}^*\mathbf{v}_j+\sum_{i=1}^{N}e^{j\theta_i}h_{k,i}\mathbf{v}_i^H\mathbf{p}_k\nonumber\\
&+\sum_{j=1}^{N} e^{-j\theta_j}h_{k,j}^*\mathbf{p}_k^H\mathbf{v}_j+\mathbf{t}_k^{H}\mathbf{Q}\mathbf{t}_k\nonumber\\
=&\sum_{n=1}^{N}\left|h_{k,n}\right|^2\mathbf{v}_n^H\mathbf{v}_n+\sum_{i=1}^{N}\sum_{j=1,j\neq i}^{N} e^{j(\theta_i-\theta_j)}h_{k,i}h_{k,j}^*\mathbf{v}_i^H\mathbf{v}_j\nonumber\\
&+ \sum_{i=1}^{N}2c_{i,k}\cos\left(\theta_i+\omega_{i,k}\right)+\mathbf{t}_k^{H}\mathbf{Q}\mathbf{t}_k\nonumber\\
=&a_k+\sum_{i=1}^{N}\sum_{j=i+1}^{N}2b_{i,j,k} \cos(\theta_i-\theta_j+\psi_{i,j,k}) \nonumber\\
&+\sum_{i=1}^{N}2c_{i,k}\cos\left(\theta_i+\omega_{i,k}\right)\label{condition},
\end{align}
where $a_k=\sum_{n=1}^{N}\left|h_{k,n}\right|^2\mathbf{v}_n^H\mathbf{v}_n+\mathbf{t}_k^{H}\mathbf{Q}\mathbf{t}_k$, $b_{i,j,k}=\left|h_{k,i}h_{k,j}^*\mathbf{v}_i^H\mathbf{v}_j\right|$, $c_{i,k}=\left|t_{k,i}\mathbf{v}_i^H\mathbf{p}_k\right|$, $\psi_{i,j,k}=\arg\left(h_{k,i}h_{k,j}^*\mathbf{v}_i^H\mathbf{v}_j\right)$, $\omega_{i,k}=\arg\left(h_{k,i}\mathbf{v}_i^H\mathbf{p}_k\right)$ are  constants.
From (\ref{condition}), it follows that the condition (\ref{constraint_Problem3}) in Problem (P2.2) can be simplified as
 \begin{align}
\gamma\leq& a_k+\sum_{i=1}^{N}\sum_{j=i+1}^{N}2b_{i,j,k} \cos(\theta_i-\theta_j+\psi_{i,j,k})\nonumber\\
&+\sum_{i=1}^{N}2c_{i,k}\cos\left(\theta_i+\omega_{i,k}\right),~k=1,\cdots,K \label{condition_P2}
\end{align}
Based on (\ref{condition_P2}), we can obtain the semi-closed-form solution for Problem (P2.2) can be obtained as the following proposition.
\begin{Proposition}\label{Proposition_P22_kkt}
 The necessary conditions that the optimal solution $\left\{\boldsymbol{\theta}_{\mathbf{Q}},\gamma^*\right\}$  of Problem (P2-2) needs to satisfy are given as (\ref{condition_P2}) and
\begin{align}
\lambda_{k}\left(\gamma-a_k-\sum_{i=1}^{N}\sum_{j=i+1}^{N}2b_{i,j,k} \cos(\theta_i-\theta_j+\psi_{i,j,k})\right.\nonumber\\
\left.-\sum_{i=1}^{N}2c_{i,k}\cos\left(\theta_i+\omega_{i,k}\right)\right)=0,&\nonumber\\
\sum_{j=1,j\neq n}^{N}\bar{b}_{n,j}\sin\left(\theta_n-\theta_j+\bar{\psi}_{n,j}\right)+\bar{c}_{n}\sin\left(\theta_n+\bar{\omega}_{n}\right)=0,&\nonumber\\
n=1,\cdots,N,&\nonumber\\
-1+\sum_{k=1}^{K}\lambda_{k}=0,&\nonumber\\
k=1,\cdots,K,&
\label{KKT_P2-2}
\end{align}
where $\bar{b}_{n,j}=\sum_{k=1}^{K}2\lambda_{k}b_{n,j,k}$,  $\bar{c}_{n}=\sum_{k=1}^{K}2\lambda_{k}c_{n,k}$, $\bar{\psi}_{n,j}$ and $\bar{\omega}_{n}$ are the auxiliary angle of $\psi_{n,j,k}$ and $\omega_{n,k}$, $k=1,\cdots,K$, respectively.
\end{Proposition}
\begin{IEEEproof}
Lagrangian function of Problem (P2.2) is given as
\begin{align}
L=-\gamma+\sum_{k=1}^{K}\lambda_{k}&\left[-\sum_{i=1}^{N}\sum_{j=i+1}^{N}2b_{i,j,k} \cos(\theta_i-\theta_j+\psi_{i,j,k})\nonumber \right.\\
&\left.~-\sum_{i=1}^{N}2c_{i,k}\cos\left(\theta_i+\omega_{i,k}\right)+\gamma-a_k\right].
\label{Lagran}
\end{align}
Based on \cite{IEEEhowto:S.Boyd},  we obtain that for the non-convex problem, by taking the derivative of the Lagrangian function (\ref{Lagran}),
we can obtain the KKT optimality conditions in (\ref{KKT_P2-2}), which are the necessary conditions that the optimal solution  $\left\{\boldsymbol{\theta}_{\mathbf{Q}},\gamma^*\right\}$ needs to satisfy.
\end{IEEEproof}

We denote the solution set of equation set (\ref{KKT_P2-2})  as $\Omega$, which can be obtained by the interval iterative method \cite{ortega1970iterative}.  The main ideas of this method are given as follow: First, we divide the domain of $\boldsymbol{\theta}$ into a finite number of initial intervals and make sure that each interval has a unique solution (When the radius of the interval is greater than the radius of its Krawczyk operator, the interval has a unique solution). Then, when we make sure each interval only has one solution, we start to obtain the solution via the iteration method. In each iteration, we obtain the intersection of the intervals and the its Krawczyk operator as the new interval, and then compute the new Krawczyk operator of the new interval for the next iteration. Until the radius of the interval is smaller than an error.  The center of the interval can be regarded as a solution.

 From (\ref{Problem_2-2}), it is observed that the optimal solution $\left\{\boldsymbol{\theta}_{\mathbf{Q}},\gamma^*\right\}$ can be determined by the maximums of all $\gamma$, which belong to set $\Omega$, i.e.,
\begin{align}
\gamma^*=\max\left\{\gamma: \left\{\boldsymbol{\theta},\gamma\right\}\in \Omega  \right\},\label{optimal_tha}
\end{align}
and the corresponding $\boldsymbol{\theta}_{\mathbf{Q}}$ is optimal for Problem (P2.2).

\subsubsection{Algorithm Summary}
The detailed alternating optimization for Problem (P2) is summarized as Algorithm \ref{table_AO}. In each iteration, we first optimize $\mathbf{Q}_{i+1}$ with the fixed $\boldsymbol{\theta}_{i}$, which is obtained at the previous iteration. Then,  we optimize $\boldsymbol{\theta}_{i+1}$ for the fixed $\mathbf{Q}_{i+1}$ by Proposition  \ref{Proposition_P22_kkt} and (\ref{optimal_tha}),  which is obtained at the current iteration.
\begin{algorithm}
        \caption{ Alternating Optimization for Problem (P2) }
\label{table_AO}
        \begin{algorithmic}[1]
\Require $\mathbf{G}_k$, $k=1,\cdots, K$, $\mathbf{H}$, $P_{\max}$,  and error tolerances $\delta>0$.
 \Ensure $\left\{\mathbf{Q}_{*},\boldsymbol{\theta}_{*}\right\}$.
 \State   {\bf Initialize}  Randomly generate $J$ independent realizations of  $\boldsymbol{\theta}^{(j)}$, $j=1,\cdots,J$, and compute the  corresponding optimal $\mathbf{Q}^{(j)}$, $j=1,\cdots,J$, according CVX. Select initial value $\left\{\mathbf{Q}_1,\boldsymbol{\theta}_1\right\}$ as the realization yielding the largest objective value of Problem (P2)
\While{$\left|\gamma_i-\gamma_{i-1}\right|>\delta$}  as
\State Compute $\mathbf{Q}_{i+1}$ with the fixed $\boldsymbol{\theta}_{i}$  by CVX.
\State Compute $\boldsymbol{\theta}_{i+1}$ with the fixed $\mathbf{Q}_{i+1}$ by Proposition  \ref{Proposition_P22_kkt} and (\ref{optimal_tha}).
\State Let $\gamma_{i+1}=\max\left\{\mathbf{u}^{H}_{i+1}\mathbf{G}_k  \mathbf{H}\mathbf{Q}_{i+1}\mathbf{H}^{H}\mathbf{G}_k^{H}\mathbf{u}_{i+1}\right\}$
\EndWhile
\State Let $\left\{\mathbf{Q}_{*},\boldsymbol{\theta}_{*}\right\}=\left\{\mathbf{Q}_{i+1},\boldsymbol{\theta}_{i+1}\right\}$
 \end{algorithmic}
\end{algorithm}

\subsection{Optimal Solution for the Special Case}

The previous two subsections proposed two numerical algorithms to obtain the phase shifts and covariance matrix for Problem (P1), which loses insight and cannot obtain the optimal solution for Problem (P1). However, the semi-closed-form optimal solution can be obtained in a special case that we only need to maximize the capacity of a MU.   We first show the sufficient and necessary conditions that the special case happens, and then compute the semi-closed-form optimal solution of the special case.

For the convenience, Problem (P1) can be rewritten as
\begin{align}
\max_{\left\{\mathbf{Q},\boldsymbol{\theta}\right\}\in\mathcal{Q}}\min_{k=1,\cdots,K}R_k\left(\mathbf{Q},\boldsymbol{\theta}\right),
\end{align}
where $\mathcal{Q}=\left\{\mathbf{Q},\boldsymbol{\theta}\left|(\ref{SJ_QP}),(\ref{SJ_Q}),(\ref{SJ_theta})\right.\right\}$ is the feasible set for Problem (P1), and $R_k\left(\mathbf{Q},\boldsymbol{\theta}\right)$ is defined as
\begin{align}
R_k\left(\mathbf{Q},\boldsymbol{\theta}\right)=\log\left(1+\left(\mathbf{u}^{H}\mathbf{G}_{k} \mathbf{H}+\mathbf{t}_{k}^{H}\right)\mathbf{Q}\left(\mathbf{H}^{H}\mathbf{G}^{H}_{k} \mathbf{u}+\mathbf{t}_{k}\right)\right).
\end{align}

First, the following proposition shows the sufficient and necessary conditions that the special case happens.
\begin{Proposition}\label{proposition_special_case}
The optimal solution for $\max_{\left\{\mathbf{Q},\boldsymbol{\theta}\right\}\in\mathcal{Q}}R_{k_0}\left(\mathbf{Q},\boldsymbol{\theta}\right)$ is denoted as  $\left\{\mathbf{Q}_{k_0},\boldsymbol{\theta}_{k_0}\right\}$, $k_0\in\left\{1,\cdots,K\right\}$. If and only if
\begin{align}
R_{k_0}\left(\mathbf{Q}_{k_0},\boldsymbol{\theta}_{k_0}\right)\leq R_{k}\left(\mathbf{Q}_{k_0},\boldsymbol{\theta}_{k_0}\right),~k=1,\cdots,K, \label{condition_k0}
\end{align}
then $\left\{\mathbf{Q}_{k_0},\boldsymbol{\theta}_{k_0}\right\}$ is the optimal solution for Problem (P1).
\end{Proposition}
\begin{IEEEproof}
The necessary of the conditions (\ref{condition_k0}) is obvious. We only prove the sufficiency of the conditions (\ref{condition_k0}). We denote $\left\{\mathbf{Q}_{*},\boldsymbol{\theta}_{*}\right\}$ is the optimal solution for Problem (P1). Due to (\ref{condition_k0}), it follows
\begin{align}
\min_{k=1,\cdots,K}R_k\left(\mathbf{Q}_{*},\boldsymbol{\theta}_{*}\right)\leq & R_{k_0}\left(\mathbf{Q}_{*},\boldsymbol{\theta}_{*}\right)\nonumber\\
\leq & R_{k_0}\left(\mathbf{Q}_{k_0},\boldsymbol{\theta}_{k_0}\right)\nonumber\\
\leq & R_{k}\left(\mathbf{Q}_{k_0},\boldsymbol{\theta}_{k_0}\right) ,~k=1,\cdots,K. \label{proof_condtion_k0}
\end{align}
Based on (\ref{proof_condtion_k0}), we obtain $\min_{k=1,\cdots, K}R_k\left(\mathbf{Q}_{*},\boldsymbol{\theta}_{*}\right)\leq  \min_{k=1,\cdots,K}R_k\left(\mathbf{Q}_{k_0},\boldsymbol{\theta}_{k_0}\right)$,  which implies that  $\left\{\mathbf{Q}_{k_0},\boldsymbol{\theta}_{k_0}\right\}$ is the optimal solution for Problem (P1).
\end{IEEEproof}

\begin{Remark}
From Proposition \ref{proposition_special_case}, we observe that in this special case, even if we only maximize the capacity of the MU $k_0$, its capacity is still smaller than the capacities of other MUs. In other words, in this special case, even if we optimize channel condition for MU $k_0$ by designing the phase shift $\boldsymbol{\theta}$, the MU $k_0$ still own the worst channel condition. It implies that the multicast capacity only depends on the capacity for the MU $k_0$.
\end{Remark}

Next, we show how to obtain $\left\{\mathbf{Q}_{k_0},\boldsymbol{\theta}_{k_0}\right\}$ by optimizing
\begin{align}
\max_{\left\{\mathbf{Q},\boldsymbol{\theta}\right\}\in\mathcal{Q}}\log\left(1+\left(\mathbf{u}^{H}\mathbf{G}_{k_0} \mathbf{H}+\mathbf{t}_{k_0}^{H}\right)\mathbf{Q}\left(\mathbf{H}^{H}\mathbf{G}^{H}_{k_0} \mathbf{u}+\mathbf{t}_{k_0}\right)\right),
\end{align}
which is equivalent to the capacity of multiple-input single-output (MISO) channel \cite{4036286}.  For the MISO channel, the capacity equals that of a single-input single-output channel with the signal transmitted  over the multiple-antenna coherently combined to maximize the channel signal noise ratio (SNR). Thus, for a fixed $\boldsymbol{\theta}$, the corresponding capacity and optimal covariance matrix $\mathbf{Q}_{\boldsymbol{\theta}}$ are respectively given as \cite{4036286}
\begin{align}
&\max_{\mathbf{Q}\in\mathcal{Q}}\log\left(1+\left(\mathbf{u}^{H}\mathbf{G}_{k_0} \mathbf{H}+\mathbf{t}_{k_0}^{H}\right)\mathbf{Q}\left(\mathbf{H}^{H}\mathbf{G}^{H}_{k_0} \mathbf{u}+\mathbf{t}_{k_0}\right) \right)\nonumber\\
=& \log\left(1+P_{\max}\left \|\left(\mathbf{u}^{H}\mathbf{G}_{k_0} \mathbf{H}+\mathbf{t}_{k_0}^{H}\right)\right \|^2 \right),\label{Problem_simo_1}\\
\mathbf{Q}_{\boldsymbol{\theta}}=&\mathbf{V}_{\boldsymbol{\theta}}^{H}\mathrm{diag}\left[P_{\max},0,\cdots, 0 \right]\mathbf{V}_{\boldsymbol{\theta}} \label{optimal_Q_case1},
\end{align}
where $\mathbf{V_{\boldsymbol{\theta}}\in \mathbb{C}^{N \times N} }$ is obtained by the singular value decomposition of  $\mathbf{u}^{H}\mathbf{G}_{k_0} \mathbf{H}+\mathbf{t}_{k_0}^{H}$, i.e.,
\begin{align}
\mathbf{u}^{H}\mathbf{G}_{k_0} \mathbf{H}+\mathbf{t}_{k_0}^{H}\triangleq \mathbf{U}_{\boldsymbol{\theta}}\Sigma_{\boldsymbol{\theta}} \mathbf{V}_{\boldsymbol{\theta}}^{H},\label{SVD}
\end{align}
where $\mathbf{U}_{\boldsymbol{\theta}}=1$, $\mathbf{V}_{\boldsymbol{\theta}}^H$ is unitary matrix, and $\Sigma_{\boldsymbol{\theta}}\in \mathbb{C}^{1 \times N} $ is a rectangular vector whose first element are non-negative real numbers and whose other elements are zero.
\begin{Remark}
From (\ref{Problem_simo_1}),  (\ref{optimal_Q_case1}) and (\ref{SVD}), it is observed that we only need to compute $\boldsymbol{\theta}_{0 }$ for maximizing $\log\left(1+P_{\max}\left \|\left(\mathbf{u}^{H}\mathbf{G}_{k_0} \mathbf{H}+\mathbf{t}_{k_0}^{H}\right)\right \|^2 \right)$. After  $\boldsymbol{\theta}_{0 }$ is obtained,  $\mathbf{Q}_{k_0}$ can be obtained by (\ref{optimal_Q_case1}) and (\ref{SVD}).
\end{Remark}

Thus, we only need to solve the following problem
\begin{align}
(\text{P3})\max_{\boldsymbol{\theta}}\left \| \left(\mathbf{u}^{H}\mathbf{G}_{k_0} \mathbf{H}+\mathbf{t}_{k_0}^{H}\right)\right \|^2.
\end{align}

Then, the optimal solution $\boldsymbol{\theta}_{k_0}$ for Problem (P3) is summarized in the following proposition.
\begin{Proposition}\label{Proposition_case1_kkt}
 The necessary conditions for the optimal solution  $\boldsymbol{\theta}_{k_0}=\left[\theta_{{k_0},1},\cdots,\theta_{{k_0},N}\right ]$ for Problem (P3)  are given as
 \begin{align}
 0=&\sum_{m=1}^{M}\sum_{j=1,j\neq n}^{N}2G_{n,j,m}^{(2)}\sin\left(\theta_{{k_0},n}-\theta_{{k_0},j}+\vartheta_{n,j,m}^{(2)}\right)\nonumber\\
&+\sum_{m=1}^{M}2B^{\left(2\right)}_{n,m}\sin \left(\theta_{{k_0},n}+\iota_{n,m}\right),\nonumber\\
n=&1,\cdots,N,  \label{KKT_ProblemP3}
 \end{align}
where $G_{n,j,m}^{(2)}=a_{h_{2,n}}a_{H_{n,m}}a_{h_{2,j}}a_{H_{j,m}}$, $\vartheta_{n,j,m}^{(2)} =\theta_{h_{2,n}}+\theta_{H_{n,m}}-\theta_{h_{2,j}}-\theta_{H_{j,m}}$, $B^{\left(2\right)}_{n,m}=a_{h_{2,n}}a_{H_{n,m}}a_{t_{2,m}}$, and $\iota_{n,m}=\theta_{h_{2,n}}+\theta_{H_{n,m}}+\theta_{t_{2,m}}$.
\end{Proposition}
\begin{IEEEproof}
Same as Proposition \ref{Proposition_P22_kkt}, and hence omitted for simplicity.
\end{IEEEproof}
 \begin{Remark}
From Proposition \ref{Proposition_case1_kkt}, we have
\begin{itemize}
\item The solution set $\mathcal{Q}$ which satisfies condition (\ref{KKT_ProblemP3}) can be obtained by  the interval iterative method \cite{ortega1970iterative}.  The optimal solution $\boldsymbol{\theta}_{k_0}$ can be determined by the maximums of all $\Upsilon \left(\boldsymbol{\theta}\right)$, where $\boldsymbol{\theta}$ belongs to set  $\mathcal{Q}$, i.e.,
\begin{align}
\Upsilon^* \left(\boldsymbol{\theta}_{k_0}\right)=\max\left\{\Upsilon \left(\boldsymbol{\theta}\right):\boldsymbol{\theta}\in \mathcal{Q}  \right\},\label{optimal_tha_1}
\end{align}
where
\begin{align}
\Upsilon \left(\boldsymbol{\theta}\right)=&\sum_{m=1}^{M}\sum_{i=1}^{N}\sum_{j=i+1}^{N}2G_{i,j,m}^{(2)}\cos \left(\theta_{i}-\theta_{j}+\vartheta_{i,j,m}^{(2)}\right)\nonumber\\
&+\sum_{m=1}^{M}\sum_{n=1}^{N}2B^{\left(2\right)}_{n,m}\cos \left(\theta_{n}+\iota_{n,m}\right).\label{function_UP}
\end{align}
If the special case happens, we adopt (\ref{optimal_tha_1})  to obtain the optimal solution; else we adopt the gradient descent method and alternating optimization to solve Problem (P2).

\item  Proposition \ref{Proposition_case1_kkt} only presents the semi-closed-form optimal solution for Problem (P3). Here, we set $M=1$ for example to demonstrate Proposition \ref{Proposition_case1_kkt} for (P3), which has the closed-form solutions.

When $M=1$,  (P3)  can be rewritten as
\begin{align}
&\max_{\boldsymbol{\theta}}\sum_{i=1}^{N}\sum_{j=i+1}^{N}2G_{i,j,1}^{(2)}\cos \left(\theta_{{k_0},i}-\theta_{{k_0},j}+\vartheta_{i,j,1}^{(2)}\right)\nonumber\\
+&\sum_{i=1}^{N}2B^{\left(2\right)}_{i,1}\cos \left(\theta_{{k_0},1}+\iota_{i,1}\right).\label{Remak_case1_2}
\end{align}
Therefore, $\theta_{{k_0},n}=-\left(\theta_{h_{2,n}}+\theta_{H_{n,1}}+\theta_{t_{2,1}}\right)$, $n=1,\cdots, N$, is optimal for (\ref{Remak_case1_2}), which is the optimal solution for $M=1$ as well.
\end{itemize}
\end{Remark}

\begin{figure}[t!]
\centering\includegraphics[width=3.2in]{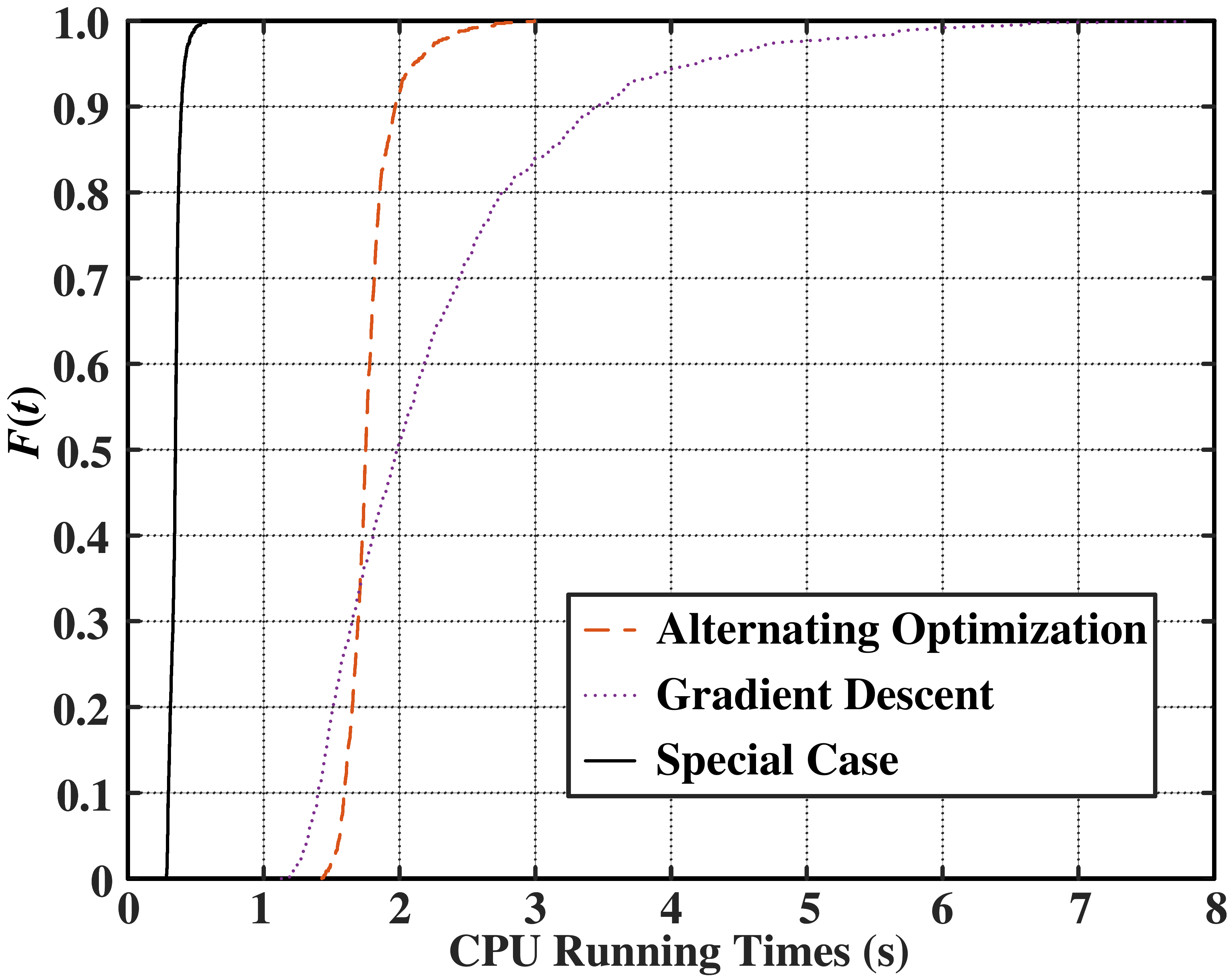}
\caption{The cumulative distribution function of running times. }\label{fig_iterations}
\end{figure}
Fig. \ref{fig_iterations} shows the cumulative distribution function of running times under the gradient descent, the alternating optimization, and the special case, where $\rho=20$ (dB),  $N=8$, and $M=8$.  It is observed that the running time of the special case is smaller than those of gradient descent and the alternating optimization.

\section{Asymptotic Analysis}\label{section_asymptotic}
In the previous two sections, the capacity of RIS assisted  multicast transmission is maximized by the numerical algorithms, which lose some intuition for the performance of RIS in multicast transmission. Since the reflecting elements are low-cost with a simple structure, the RIS can integrate a large number of the reflecting elements \cite{8319526}. Thus, it is worth to study the asymptotic behaviors of  capacity $C$  when some or all of the number of reflecting elements, BS antennas, and MUs go to infinity.    We consider  $\mathbf{H}$, $\mathbf{G}_{k}^{H}$, and $\mathbf{t}_{k}^{H}$ as Rician Fading. The channels of  the BS-RIS link, the RIS-MUs links, and the BS-MUs links  are denoted as $\mathbf{H}=\sqrt{\frac{B}{B+1}}\bar{\mathbf{H}}+\sqrt{\frac{1}{B+1}}\hat{\mathbf{H}}$, $\mathbf{G}_k=\sqrt{\frac{B}{B+1}}\text{diag}\left(\bar{\mathbf{h}}_k\right)+\sqrt{\frac{1}{B+1}}\text{diag}\left(\hat{\mathbf{h}}_k\right)$, and $\mathbf{t}_k=\sqrt{\frac{B}{B+1}}\bar{\mathbf{t}}_k+\sqrt{\frac{1}{B+1}}\hat{\mathbf{t}}_k$, respectively.
Here, $B$ is the Rician factor; $\hat{\mathbf{H}}$, $\hat{\mathbf{h}}_k$, and $\hat{\mathbf{t}}_k$ are the non-line-of-sight (NLoS) component, and elements of  $\hat{\mathbf{H}}$, $\hat{\mathbf{h}}_k$, and $\hat{\mathbf{t}}_k$  are i.i.d., complex Gaussian distribution  $\mathcal{CN}\left(0,1\right)$.
The line-of-sight  (LoS) components are expressed by uniform rectangular array (URA). The array response of URA is given as
$\mathbf{a}_M^{H}\left(\omega,\vartheta\right)=\text{vec}\left(\mathbf{a}_{M_1}\left(\omega,\vartheta\right)\mathbf{a}_{M_2}^{H}\left(\omega,\vartheta\right)\right)$,
where $\mathbf{a}_{M_1}=\left[1,e^{j2\pi \frac{d}{\lambda} \sin \omega \cos \vartheta },\cdots,e^{j2\pi \frac{d}{\lambda}\left(M_1-1\right) \sin \omega \cos \vartheta }\right]$, $\mathbf{a}_{M_2}=\left[1,e^{j2\pi \frac{d}{\lambda} \sin \omega \sin \vartheta },\cdots,e^{j2\pi \frac{d}{\lambda}\left(M_1-1\right) \sin \omega \sin \vartheta }\right]$; $\omega$ and $\vartheta$ are the angle of departure (AoD) or angler of arrival (AoA) for x-axis and y-axis.
Thus, the LoS components $\mathbf{H}$, $\mathbf{h}_{k}^{H}$, and $\mathbf{t}_{k}^{H}$ are expressed as $\bar{\mathbf{H}}=\mathbf{a}_N\left(\omega_{AoA},\vartheta_{AoA}\right)\mathbf{a}_M^{H}\left(\omega_{AoD},\vartheta_{AoD}\right)$, $\bar{\mathbf{h}}_k=\mathbf{a}_N\left(\omega_{AoD},\vartheta_{AoD}\right)$, and $\bar{\mathbf{t}}_k=\mathbf{a}_M\left(\omega_{AoD},\vartheta_{AoD}\right)$,
respectively, where $\omega_{AoA}$ and $\vartheta_{AoD}$ are  AoA, and $\omega_{AoA}$ and  $\vartheta_{AoA}$  are the AoD, respectively; $d$ and $\lambda$ are the antenna separation and wavelength, respectively.

\subsection{Fixed MUs, Increasing Antennas and Reflecting Elements}
First, we consider the asymptotic behaviors of the maximal capacity $C$ at the cases where the number of MUs is fixed while either the number of reflecting elements or that of  BS antennas goes to infinity.
\begin{Proposition}\label{Proposition_asymptotic_MN}
If $K$ is fixed, the order growth of $C $ is given as follows.
\begin{itemize}
\item When $N$ is fixed and $M$ goes to infinity,  $C$  grows at the following rate
\begin{align}
C \approx \mathcal{O}\left(\log M\right). \label{Asymptotic_M}
\end{align}
\item When $M$ is fixed and $N$ goes to infinity, $C$  grows at the following rate
\begin{align}
C \approx \mathcal{O}\left(\log N^2\right). \label{Asymptotic_N}
\end{align}
\item When both of $N$ and $M$ go to infinity,  $C$  grows at the following rate
\begin{align}
C \approx \mathcal{O}\left(\log \left(N^2M\right)\right). \label{Asymptotic_NM}
\end{align}
\end{itemize}
\end{Proposition}
\begin{IEEEproof}
Please see Appendix \ref{proof_asymptotic_MN}.
\end{IEEEproof}
\begin{Remark} \label{Remark_asymptotic_MN}
From Proposition \ref{Proposition_asymptotic_MN}, we observe
\begin{itemize}
\item $C$ grows logarithmically with $N^2$ and $M$, as either the number of reflecting elements or that of BS antennas goes to infinity.  It follows that the increase in the numbers of antennas and reflecting elements can improve the capacity.  This is due to the fact that the diversity order increases as $N$ and $M$. Moreover, the increase in $N$ can provide more performance gain than that of $M$. This is due to the fact that with increasing of $N$, the RIS can provide more the transmit diversity and the receive diversity. The simulations in  Fig. \ref{fig_Nmax}  also confirms the results in Proposition \ref{Proposition_asymptotic_MN}.

\item
From the results of Proposition \ref{Proposition_asymptotic_MN}, we observed that $C$ goes to infinity.  In reality, $C$ will converge to a finite value even if $N$ and $M$ go to infinity since the power budget is limited. The main reason for this result is that the Rician channel model assumption is ideal. For Proposition 4.1, we focus on observing the tendency of capacity.

\item From the proof of Proposition \ref{Proposition_asymptotic_MN}, it is obtained that  $C \approx \mathcal{O}\left(\log \left(\frac{B}{B+1}\right)^2\right)$, when $B$ goes to infinity.
\item The results in Proposition \ref{Proposition_asymptotic_MN} are under the i.i.d. channel coefficients. However, in some scenarios, the channel coefficients may not be i.i.d., and it leads that the covariances between $g_{k,n,m}$ and $\bar{g}_{k,i,m}$ are not zero. Thus, in the proof of Proposition \ref{Proposition_asymptotic_MN}, (\ref{proof_hH_3}) is rewritten as
\begin{align}
\mathbb{E}\left\{ \left|\Psi_{m,k}\right|^2\right\}&=A\left(N^2\right)+\sum_{n=1}^{N}\sum_{i=1,i\neq n}^{N}\mathbf{Cov}\left(g_{k,n,m},\bar{g}_{k,i,m}\right)\nonumber\\
&\leq A\left(N^2\right)+N\left(N-1\right)\left(1-\left(\frac{B}{B+1}\right)^2\right).
\end{align}
By the same idea as proof of Proposition \ref{Proposition_asymptotic_MN} and Birkhoff's ergodic theory, we obtain that (\ref{Asymptotic_M}), (\ref{Asymptotic_N}), and (\ref{Asymptotic_NM}) are rewritten as
\begin{align}
&C \leq \mathcal{O}\left(\log N^2\right)\\
&C \leq \mathcal{O}\left(\log M\right)\\
&C \leq \mathcal{O}\left(\log \left(N^2M\right)\right),
\end{align}
respectively.
\end{itemize}
\end{Remark}
\subsection{Increasing MUs, Fixed Antennas and Reflecting Elements}
Then, we consider the asymptotic behaviors of the maximal capacity $C$ at the case where the number of MUs $K$ goes to infinity while both of $N$ and $M$ are fixed. We can see that $\min_{k=1,\cdots,K}\left\{\left\|\mathbf{h}_{k}^{H}\right\|\right\}\rightarrow0$ and $\min_{k=1,\cdots,K}\left\{\left\|\mathbf{t}_{k}^{H}\right\|\right\}\rightarrow0$, as $K\rightarrow\infty$, which implies that at least one RIS-MU link and at least one BS-MU link both being completely unavailable,  and we also have $C \rightarrow0$, which is discussed in detail as follows.
\begin{Proposition}\label{Proposition_asymptotic_K}
When both of $N$ and $M$ are fixed, and $K$ goes to infinity, the maximal capacity $C$ goes to $0$ at the following rate
\begin{align}
C \approx \mathcal{O}\left(\frac{1}{K^{1/\left(N^2M\right)}}\right).
\end{align}
\end{Proposition}
\begin{IEEEproof}
From the proof of Proposition \ref{Proposition_asymptotic_MN}, we obtain that $\left\|\mathbf{u}^{H}\mathbf{G}_k \mathbf{H}+\mathbf{t}_{k}^{H}\right\|^2$ follows the non-central chi-square distribution with a mean of $MA\left(N^2\right)$ and $2M$ degrees of freedom.
It is concluded that the minimum of $K$ non-central chi-squared random variables $\left\|\mathbf{u}^{H}\mathbf{G}_k \mathbf{H}+\mathbf{t}_{k}^{H}\right \|^2$, i.e., $\min_{k=1,\cdots,K}\left\{\left\|\mathbf{u}^{H}\mathbf{G}_k \mathbf{H}+\mathbf{t}_{k}^{H}\right \|^2\right\}$, can be scaled as $K^{-1/\left(MA\left(N^2\right)\right)}$ \cite{4036286}.

Note that  $C $ is upper bounded by the minimum of the point-to-point capacity of the RIS system as:
\begin{align}
C &\leq\log\left(1+P_{\max}\min_{k=1,\cdots,K}\left \|  \mathbf{u}^{H}\mathbf{G}_k \mathbf{H}+\mathbf{t}_{k}^{H}\right \|^2 \right)\nonumber\\
&\approx \log\left(1+\frac{P_{\max}}{K^{1/\left(MA\left(N^2\right)\right)}}\right)\nonumber\\
&\approx  \frac{P_{\max}}{K^{1/\left(MA\left(N^2\right)\right)}},
\end{align}
which is $\mathcal{O}\left(\frac{1}{K^{1/N^2M}}\right)$, and $C$ is lower bounded by the spatially white rate \cite{4036286}:
\begin{align}
C &\geq \log\left(1+\frac{P_{\max}}{N}\min_{k=1,\cdots,K}\left \|  \mathbf{u}^{H}\mathbf{G}_k \mathbf{H}+\mathbf{t}_{k}^{H}\right \|^2 \right)\nonumber\\
&\approx  \frac{P_{\max}}{NK^{1/\left(MA\left(N^2\right)\right)}},\label{lower_bound_K}
\end{align}
which is also $\mathcal{O}\left(\frac{1}{K^{1/N^2M}}\right)$.
\end{IEEEproof}
From Proposition \ref{Proposition_asymptotic_K},  it is obtained that $C$ is the inverse proportion with the number of MUs $K$.
The declining ratio decreases with the increasing $M$ and $N$. It implies that the increase in the $M$ and $N$ can reduce the negative effect of the number of MUs. This result in Proposition \ref{Proposition_asymptotic_K} has been confirmed in Fig. \ref{fig_Kmax}.

\subsection{Increasing MUs, Antennas and Reflecting Elements}
Finally, we consider the asymptotic behaviors of the maximal capacity $C$ at the case where all of $N$, $M$ and $K$ go to infinity.
\begin{Proposition}\label{Proposition_asymptotic_KMN}
When both $M$ and $K$ go to infinity at the ratio $0<\frac{M}{K}<\infty$, the order growth of $C$ is given as follows:
\begin{itemize}	
\item If $N$ is fixed, we have
\begin{align}
 \mathbb{E}\left\{C \right\}\approx \mathcal{O}\left(1\right).\label{Asymptotic_KM}
\end{align}
\item If $N$ goes to infinity, we have
 \begin{align}
  \mathcal{O}\left(1\right)\leq \mathbb{E}\left\{C \right\}\leq \mathcal{O}\left(\log(N^2)\right).\label{Asymptotic_KMN}
\end{align}
\end{itemize}
\end{Proposition}
\begin{IEEEproof}
See Appendix \ref{proof_asymptotic_K}.
\end{IEEEproof}
From Proposition \ref{Proposition_asymptotic_KMN}, we observe the following results. When the number of reflecting elements $N$ is fixed, $ \mathbb{E}\left\{C \right\}$ remains constant as the number of MUs and antennas are taken to infinity at a fixed ratio.  When $N$ goes to infinity,  $ \mathbb{E}\left\{C \right\}$ is lower bounded by a constant, and its growth rate is not greater than $\log N^2$.  The above results conclude that increasing the numbers of reflecting elements and antennas can effectively counter the negative effect caused by the increase in MUs. These results are verified by Fig. \ref{fig_KM}.

From the proof of Proposition \ref{Proposition_asymptotic_KMN}, we can observe that this proof is not dependent on i.i.d. channel assumption. Thus, the results in Proposition \ref{Proposition_asymptotic_KMN} are applicable to non-i.i.d. channel.

\section{Numerical Results} \label{section_Numerical}

This section presents the numerical results of  capacity by the following two algorithms: 1) the gradient descent method, 2) alternating optimization.
 The channel model is same to section \ref{section_asymptotic}.  Moreover, we set $\rho=\frac{P_{\max}}{\sigma^2}$, $M=16$, $\frac{d}{\lambda}=1$, $\omega_{AoD}$, $\omega_{AoD}$, $\vartheta_{AoA}$, and $\vartheta_{AoD}$ are randomly set within $\left[0,2\pi\right)$.
The following numerical results are averaged over 1000 random realization.
\subsection{Benchmark}
 In comparison, we compute the multicast capacity with the following scheme:
\subsubsection{Brute-force search}  The optimal solution for Problem (P1) is obtained by brute-force search.
\subsubsection{Lower Bound} The lower bounds of the capacity are obtained by asymptotic analysis.
  \subsubsection{Beamforming design} The covariance matrix $\mathbf{Q}$ degenerates to beamforming vector $\mathbf{v}$. The capacity in Problem (P1) is rewritten as
\begin{align}
\max_{\left\|\mathbf{v}\right\|^2\leq P_{\max}} \min_{k=1,\cdots K} \log\left(1+\left\|\left(\mathbf{u}^{H}\mathbf{G}_k \mathbf{H}+\mathbf{t}_{k}^{H}\right)\mathbf{v}\right\|^2\right). \label{problem_beamforing}
\end{align}
The solution is given in \cite{9076830} in detail.

\subsubsection{Beamforming Design with Imperfect CSI}\footnote{The beamforming design with imperfect CSI has been studied for the RIS-assisted broadcast transmission. However, there are not works considering the problem for the multicast transmission. This is because this solution of the problem for the multicast transmission is similar to that for the broadcast transmission. Thus, we regard the achievable rate for beamforming design with imperfect CSI as a baseline. }
Due to the imperfect cascaded CSI of BS-RIS-MU links, the cascaded channel $\mathbf{G}_k \mathbf{H}$ is represented as
\begin{align}
\mathbf{G}_k \mathbf{H}=\mathbf{\hat{C}}_k+\Delta \mathbf{C}_k,~~k=1,\cdots,K,
\end{align}
where $\mathbf{\hat{C}}_k$ is the estimated cascaded CSI of BS-RIS-MU link at the $k$-th MU, and $\Delta \mathbf{C}_k$ is the unknown cascaded CSI given as
\begin{align}
\left\|\Delta \mathbf{C}_k\right\|_{F}\leq  \epsilon_k,~~k=1,\cdots,K,\label{unkown_error}
\end{align}
where  $\epsilon_k$ is the radii of the uncertainty regions known at the BS, and we set $\epsilon_k=0.5$ in simulation.
The transmission rate maximization problem (\ref{problem_beamforing}) is equivalent to
\begin{align}
(\text{P4})~~\min_{\mathbf{v},\mathbf{u}} &\left\|\mathbf{v}\right\|^2\\
&(\ref{SJ_theta}),~(\ref{unkown_error}),\nonumber\\
&\log\left(1+\left\|\left(\mathbf{u}^{H}\mathbf{G}_k \mathbf{H}+\mathbf{t}_{k}^{H}\right)\mathbf{v}\right\|^2\right)\geq R,~~k=1\cdots,K,\label{Rate}
\end{align}
where $R$ is the target transmission rate and constrains (\ref{Rate}) are the worst-case SNR requirements for the MUs.
From \cite{BEFB:94}, we obtain that for the given optimal solution at the $i$-th $\left\{\mathbf{v}^{(i)},\mathbf{u}^{(i)}\right\}$, $\left\|\left(\mathbf{u}^{H}\mathbf{G}_k \mathbf{H}+\mathbf{t}_{k}^{H}\right)\mathbf{v}\right\|^2$ is linearly approximated by its lower bound at $\left\{\mathbf{v}^{(i)},\mathbf{u}^{(i)}\right\}$ as follows
\begin{align}
\text{vec}^T\left(\Delta \mathbf{C}_k\right)\mathbf{A}\text{vec}\left(\Delta \mathbf{C}_k^*\right)+2\text{Re}\left\{\mathbf{l}_k\text{vec}\left(\Delta \mathbf{C}_k^*\right)\right\}+\varphi_k,
\end{align}
where
\begin{align}
\mathbf{A}=&\mathbf{v}\mathbf{v}^{H,(i)}\otimes\mathbf{u}\mathbf{u}^{T,(i)}+\mathbf{v}^{(i)}\mathbf{v}^{H}\otimes\mathbf{u}^{(i)}\mathbf{u}^{T}-\mathbf{v}^{(i)}\mathbf{v}^{H,(i)}\otimes\mathbf{u}^{(i)}\mathbf{u}^{T,(i)},\\
\mathbf{l}_k=&\text{vec}\left(\mathbf{u}\left(\mathbf{t}_{k}^{H}+\mathbf{u}^{H,(i)}\mathbf{\hat{C}}_k\right)\mathbf{v}^{(i)}\mathbf{v}^{H}\right)\nonumber\\
&+\text{vec}\left(\mathbf{u}^{(i)}\left(\mathbf{t}_{k}^{H}+\mathbf{u}^{H}\mathbf{\hat{C}}_k\right)\mathbf{v}\mathbf{v}^{H,(i)}\right)\nonumber\\
&-\text{vec}\left(\mathbf{u}\left(\mathbf{t}_{k}^{H}+\mathbf{u}^{H,(i)}\mathbf{\hat{C}}_k\right)\mathbf{v}^{(i)}\mathbf{v}^{H,(i)}\right),\\
\varphi_k=&2\text{Re}\left\{\left(\mathbf{t}_{k}^{H}+\mathbf{u}^{H,(i)}\mathbf{\hat{C}}_k\right)\mathbf{v}^{(i)}\mathbf{v}^{H}\left(\mathbf{t}_{k}+\mathbf{\hat{C}}_k^{H}\mathbf{u}\right)\right\}\nonumber\\
&-\left(\mathbf{t}_{k}^{H}+\mathbf{u}^{H,(i)}\mathbf{\hat{C}}_k\right)\mathbf{v}^{(i)}\mathbf{v}^{H,(i)}\left(\mathbf{t}_{k}+\mathbf{\hat{C}}_k^{H}\mathbf{u}^{(i)}\right).
\end{align}
Thus, constraints (\ref{Rate}) can be reformulated as
\begin{align}
\text{vec}^T\left(\Delta \mathbf{C}_k\right)\mathbf{A}\text{vec}\left(\Delta \mathbf{C}_k^*\right)+2\text{Re}\left\{\mathbf{l}_k\text{vec}\left(\Delta \mathbf{C}_k^*\right)\right\}+\varphi_k\geq 2^{R}-1,\label{Rate1}
\end{align}
By general S-procedure \cite{BEFB:94}, conditions (\ref{unkown_error}) and (\ref{Rate1}) can be approximately rewritten as the linear matrix inequality, i.e.,
\begin{align}
\begin{bmatrix}
\varpi_k \mathbf{I}+\mathbf{A} &\mathbf{l}_k \\
\mathbf{l}_k^T& \varphi_k-(2^{R}-1)-\varpi_k\epsilon_k
\end{bmatrix} \succeq 0, \label{condition_new}
\end{align}
Based on (\ref{condition_new}), Problem (P4) can be approximately reformulated as
\begin{align}
(\text{P5})~~\min_{\mathbf{v},\mathbf{u},\boldsymbol{\varpi}} &\left\|\mathbf{v}\right\|^2\\
&(\ref{SJ_theta}),~(\ref{condition_new}),\nonumber
\end{align}
where $\boldsymbol{\varpi}=\left[\varpi_1,\cdots,\varpi_K\right]^T\geq0$ are slack variables.
Problem (P5) can be solved by the alternating optimization.

\textbf{Optimization of $\mathbf{v}$ with  fixed $\mathbf{u}$:} First, for the fixed $\mathbf{u}$, the subproblem is given as
\begin{align}
(\text{P5.1})~~\min_{\mathbf{v},\boldsymbol{\varpi}} &\left\|\mathbf{v}\right\|^2\\
&(\ref{SJ_theta}),~(\ref{condition_new}),\nonumber
\end{align}
which is convex problem and can be solved by SDP techniques.

\textbf{Optimization of $\mathbf{u}$ with  fixed $\mathbf{v}$:}
Then, for the fixed $\mathbf{v}$, the subproblem of $\mathbf{u}$ is a feasibility-check problem. By introducing slack variables $\beta_k$, $k=1\cdots,K$, constraint (\ref{Rate1}) is modified as
\begin{align}
&\text{vec}^T\left(\Delta \mathbf{C}_k\right)\mathbf{A}\text{vec}\left(\Delta \mathbf{C}_k^*\right)+2\text{Re}\left\{\mathbf{l}_k\text{vec}\left(\Delta \mathbf{C}_k^*\right)\right\}+\varphi_k\nonumber\\
\geq & 2^{R}-1-\beta_k.\label{Rate2}
\end{align}
Thus, the constraint (\ref{condition_new}) can be modified as
\begin{align}
\begin{bmatrix}
\varpi_k \mathbf{I}+\mathbf{A} &\mathbf{l}_k \\
\mathbf{l}_k^T& \varphi_k-(2^{R}-1)-\varpi_k\epsilon_k-\beta_k
\end{bmatrix} \succeq 0, \label{condition_new_1}
\end{align}
and the subproblem for the fixed $\mathbf{v}$ is formulated as
\begin{align}
(\text{P5.2})~~\max_{\mathbf{v},\boldsymbol{\varpi},\boldsymbol{\beta}} &\sum_{k=1}^{K}\beta_k\\
&(\ref{SJ_theta}),~(\ref{condition_new_1}),\nonumber\\
&\boldsymbol{\beta}\geq \mathbf{0}, \label{beta}
\end{align}
where $\boldsymbol{\beta}=\left[\beta_1,\cdots,\beta_K\right]$.
Problem (P5.2) is non-convex, and thus we adopt convex-concave procedure \cite{lipp2016variations}. For the fixed $\Phi_n^{(j)}$, the non-convex part of the constraint (\ref{SJ_theta}) are linearized by $\left|\Phi_n^{(j)}\right|^2-2\text{Re}\left\{\Phi_n^*\Phi_n^{(j)}\right\}\leq-1$. Thus, Problem (P5.2) is solved by a sequence of problems, i.e.,
\begin{align}
(\text{P5.2*})~~\max_{\mathbf{v},\boldsymbol{\varpi},\boldsymbol{\beta},\boldsymbol{\delta}} &\sum_{k=1}^{K}\beta_k-\iota^{(j)} \sum_{n=1}^{N}\delta_n\label{objective_P42}\\
&(\ref{SJ_theta}),~(\ref{condition_new_1})~(\ref{beta}),\nonumber\\
&\left|\Phi_n^{(j)}\right|^2-2\text{Re}\left\{\Phi_n^*\Phi_n^{(j)}\right\}\leq\delta_n-1,\\
&\left|\Phi_n\right|^2\leq1~n=1+\delta_n,~n=1,\cdots,N,\\
&\boldsymbol{\delta}\geq \mathbf{0},
\end{align}
Here, $\boldsymbol{\delta}=\left[\delta_1,\cdots,\delta_M\right]$ are slack variables imposed over the equivalent linear constraints of constraint (\ref{SJ_theta}). $\sum_{n=1}^{N}\delta_n$ is the penalty term in the objective function (\ref{objective_P42}). $\iota^{(j)}$ is a parameter to control the accuracy of the approximations. Obviously, when $\iota^{(j)}$ goes to infinity, Problems (P5.2*) and (P5.2) become same. It is notice that Problem (P5.2*) is convex, which can be solved by CVX. Based on above discussions, the alternating optimization for Problem(P4) is summarized as Algorithm \ref{table_P3}.
\begin{algorithm}
        \caption{Alternating Optimization for Problem (P4) }
\label{table_P3}
        \begin{algorithmic}[1]
\Require  $\mathbf{G}_k$, $k=1,\cdots, K$, $\mathbf{H}$, $R$, $\iota^{0}$, $\eta>1$,  and the error tolerances $\delta_1 >0$ and $\delta_2>0$.
 \Ensure $\left\{\mathbf{v}_{*},\boldsymbol{u}_{*}\right\}$.
             \State {\bf Initialize} $\left\{\mathbf{v}^{(i)},\boldsymbol{u}^{(i)}\right\}$ and $\boldsymbol{u}^{(j)}$, which represent the output of the $i$-th  outer iteration and the input of the $j$-th  inner iteration, respectively.
 \While{$\left|\left\|\mathbf{v}^{(i)}\right\|-\left\|\mathbf{v}^{(i-1)}\right\|\right|>\delta_1 $}
 \State Compute $\mathbf{v}^{(i+1)}$ by solving Problem (P5.1) with $\mathbf{u}^{(i)}$ and $\mathbf{v}^{(i)}$.
  \State {\bf Initialize} $\iota^{(1)}=\iota^{0}$ and $\mathbf{u}^{(1)}=\mathbf{u}^{(i)}$.
\While{$\left\|\mathbf{u}^{(j)}-\mathbf{u}^{(j-1)}\right\|>\delta_2 $}
  \State Compute $\mathbf{u}^{(j+1)}$ by solving Problem (P5.2*) with $\mathbf{u}^{(j)}$, $\mathbf{v}^{(i+1)}$, and $\iota^{(j)}$.
  \State $\iota^{(j+1)}=\eta\iota^{(j)}$.
\EndWhile
\State Let $\mathbf{u}^{(i+1)}=\mathbf{u}^{(j)}$.
\EndWhile
\State Let  $\left\{\mathbf{v}_{*},\boldsymbol{u}_{*}\right\}=\left\{\mathbf{v}^{(i+1)},\boldsymbol{u}^{(i+1)}\right\}$.
  \end{algorithmic}
\end{algorithm}

\begin{Remark}
It is worth to theoretically characterize the performance gap between the perfect CSI and the imperfect CSI.  It is hard to the direct performance gap, since we cannot obtain the closed-form solution for the two cases. However, we can obtain the lower bound of the gap by comparing the lower bound of the prefect CSI case and the upper bound of the imperfect CSI case.  To our best knowledge, there are no works considering to compute the upper bound of the imperfect CSI case, which will be an interesting research problem for future study.
\end{Remark}

\subsection{Performance Comparison}

\begin{figure}[t!]
\centering\includegraphics[width=3.2in]{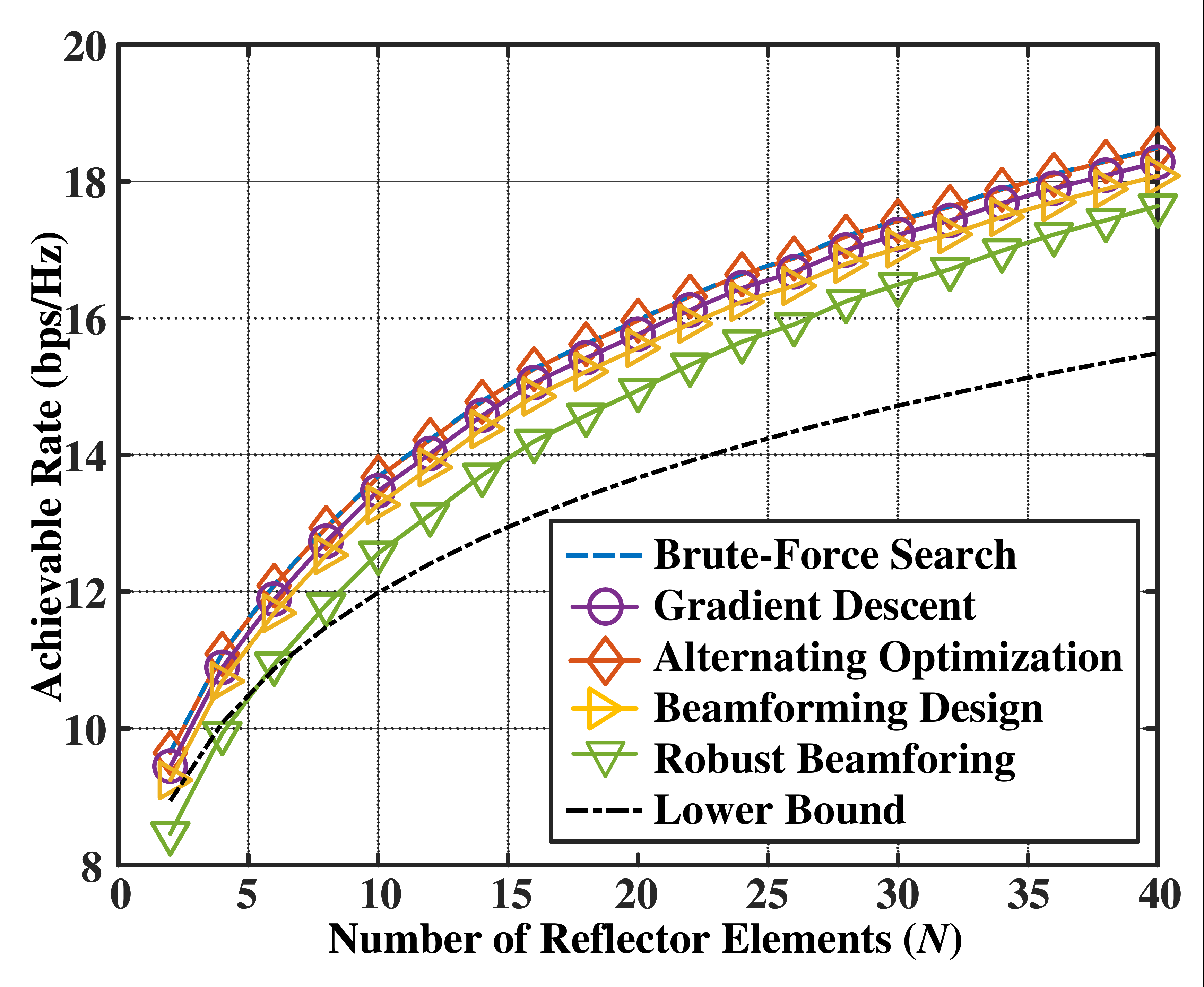}
\caption{Achievable rates  versus the number of reflecting elements ($N$).}\label{fig_Nmax}
\end{figure}
Fig. \ref{fig_Nmax} shows the achievable rates versus the number of reflecting elements $N$, where  $M=8$, $K=8$, $B=1$, and $\rho=20$ (dB).  All the schemes with RIS are better than the scheme without RIS. The achievable rate for alternating optimization closes to the maximal capacity obtained by brute-force search, which proves that the alternating optimization has a brilliant performance.  The proposed algorithms outperform the beamforming design scheme. This is because beamforming vector can be seen as one-rank covariance matrix $\mathbf{Q}$, and in most cases cannot achieves the capacity. The gaps between the curves of the proposed algorithms and the lower bound increase significantly. This is because with more reflecting elements, the RIS provides more degrees of freedom further to improve the BS-MU link with the worst channel condition and thus obtain the hight gains.

\begin{figure}[t!]
\centering\includegraphics[width=3.2in]{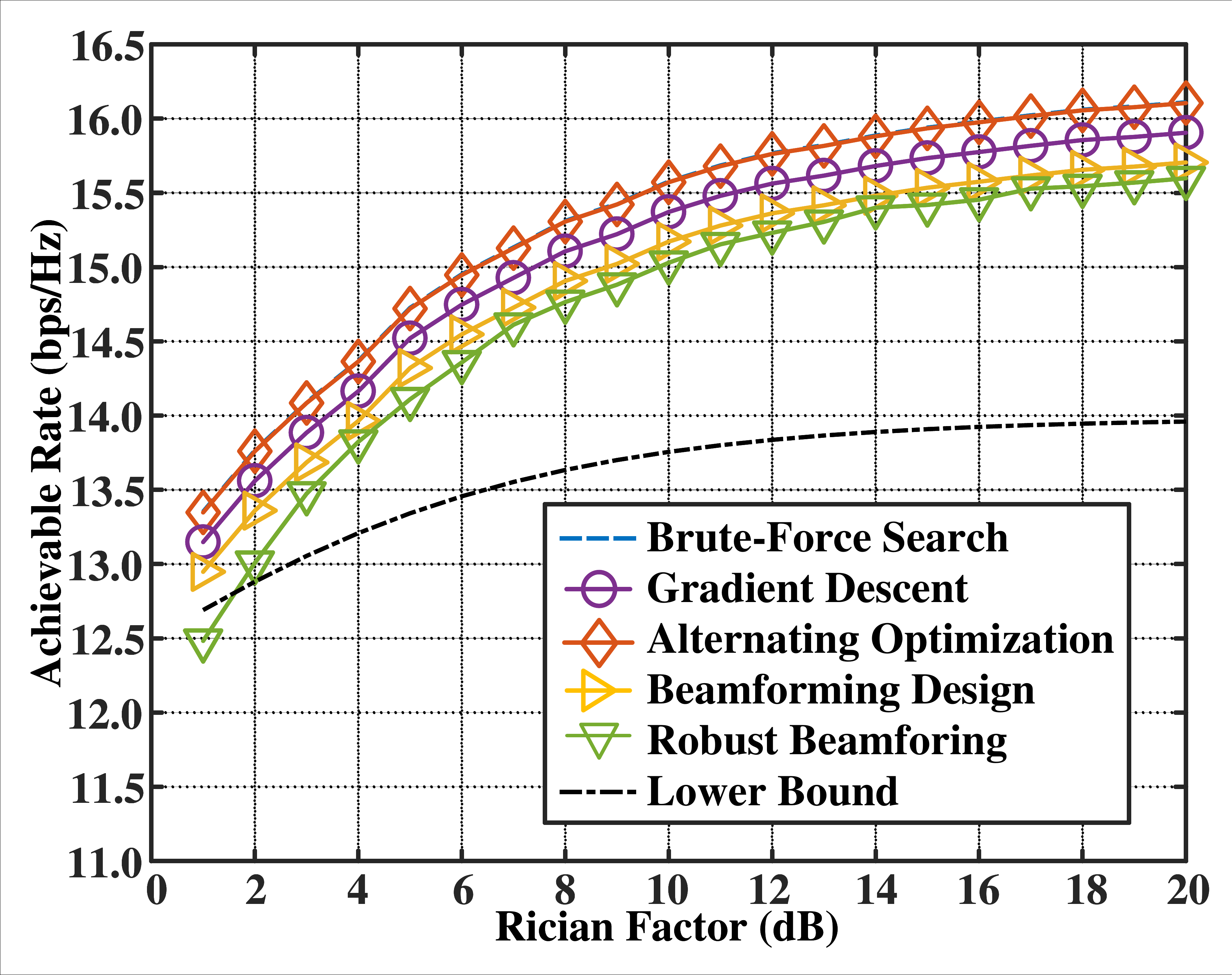}
\caption{Achievable rates versus the Rician factor ($B$).}\label{fig_Rician}
\end{figure}
Fig. \ref{fig_Rician} presents the achievable rates versus the Rician factor $B$, where $\rho=20$ (dB),  $N=8$, and $M=8$. Fig. \ref{fig_Rician} shows that the achievable rates first increase and then approach constants, which conforms Remark \ref{Remark_asymptotic_MN}.It implies that only LoS components exists, and the channel coefficients remain unchanged.  Please notice that the lower bound is the one of the capacity rather than the one of the achievable rate for the beamforming design. Thus,  the lower bound is greater than achievable rate of the beamforming design when $B\leq1.6$.

\begin{figure}[t!]
\centering\includegraphics[width=3.2in]{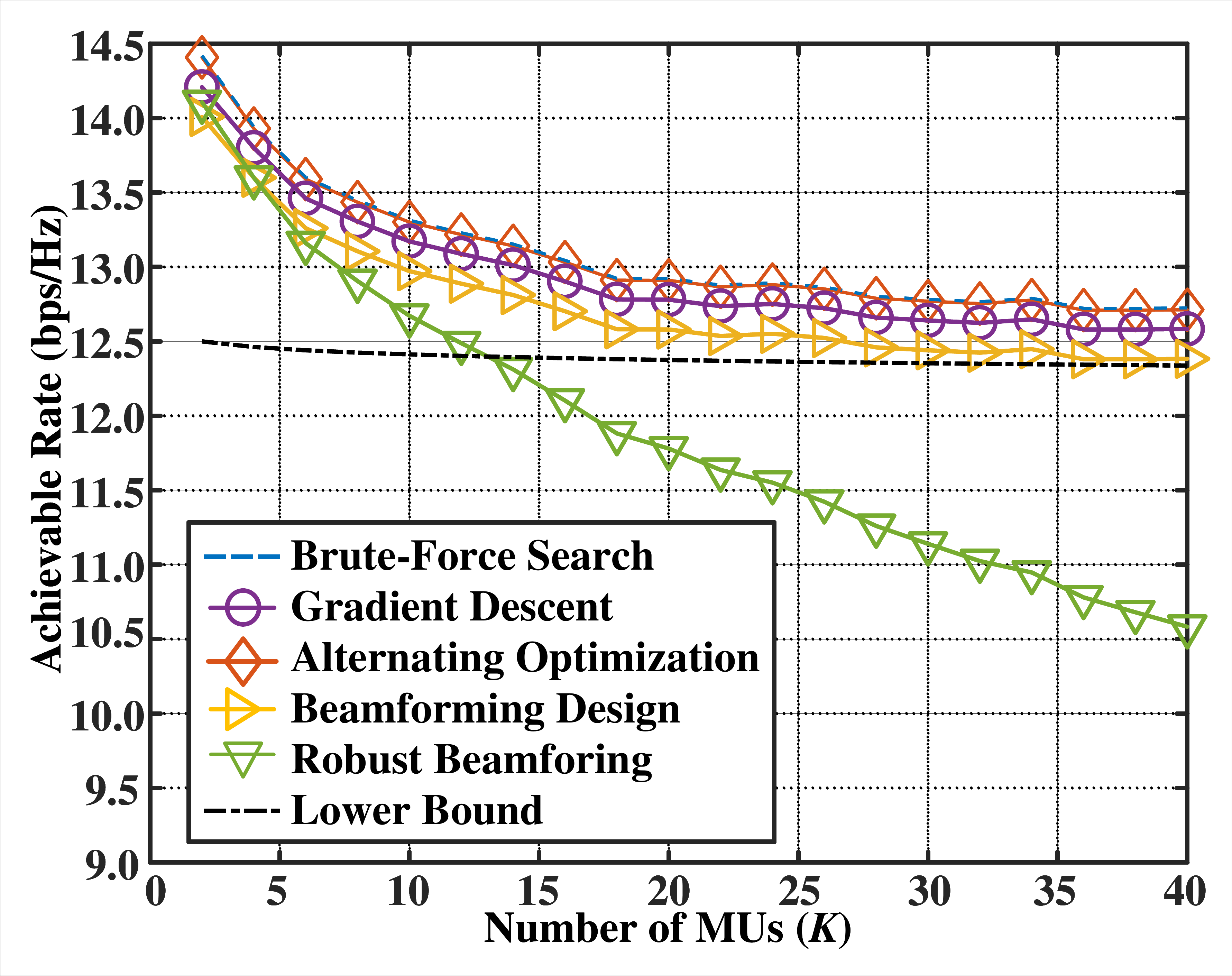}
\caption{Achievable rates versus the number of MUs ($K$).}\label{fig_Kmax}
\end{figure}
Fig. \ref{fig_Kmax} shows the relationship between the achievable rates and the number of MUs $K$, where $N=8$, $M=8$,  $B=1$, and  $\rho=20$ (dB). From Fig. \ref{fig_Kmax}, we can observe that the achievable rates decrease with the increase in $K$. Notice that the gap of the curves between the robust beamforming and the capacity increase with $K$ increasing. This is because the increase in $K$ rises the uncertainty for robust beamforming design, which leads to the reduction of the achievable rate. The lower bound is given by (\ref{lower_bound_K}), and the gap of the curves between the capacity and lower bound decrease.

 \begin{figure}[!t]
\centering\includegraphics[width=3.2in]{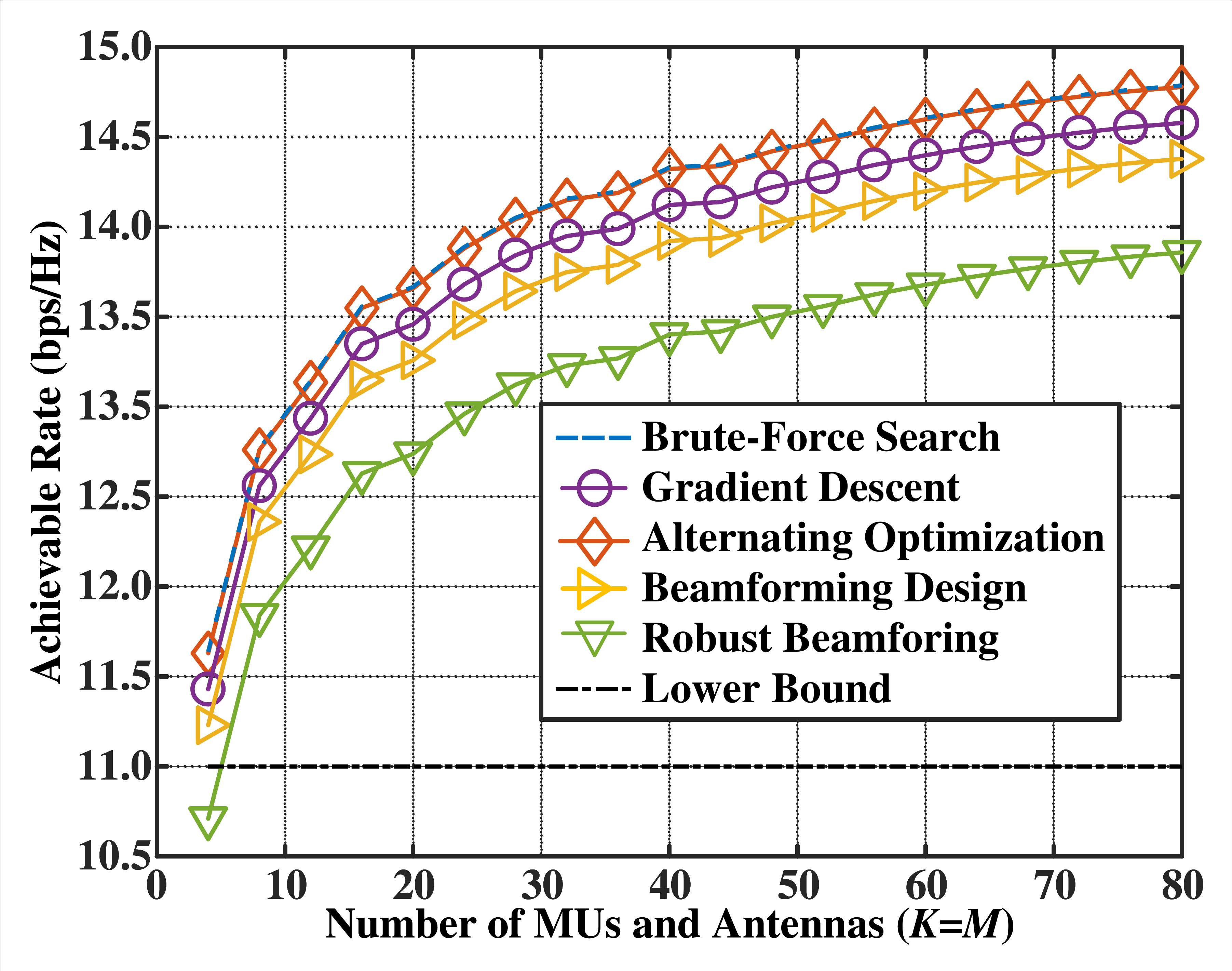}
\caption{Achievable rate versus the equal numbers of MUs ($K$) and antennas ($M$).}\label{fig_KM}
\end{figure}

 Fig. \ref{fig_KM} shows the achievable rates with identical $K$ and $M$ (i.e., $\frac{K}{M} =1$), where $N=8$ and $B=1$.  The lower bound given by (\ref{lower_bound_KM}) is a constant.  The achievable rates rise firstly and then approach a finite value. It follows that the capacity is limited when $M$ and $K$ go to infinity at a ratio, which confirms Proposition \ref{Proposition_asymptotic_KMN}.

 \begin{figure}[!t]
\centering\includegraphics[width=3.2in]{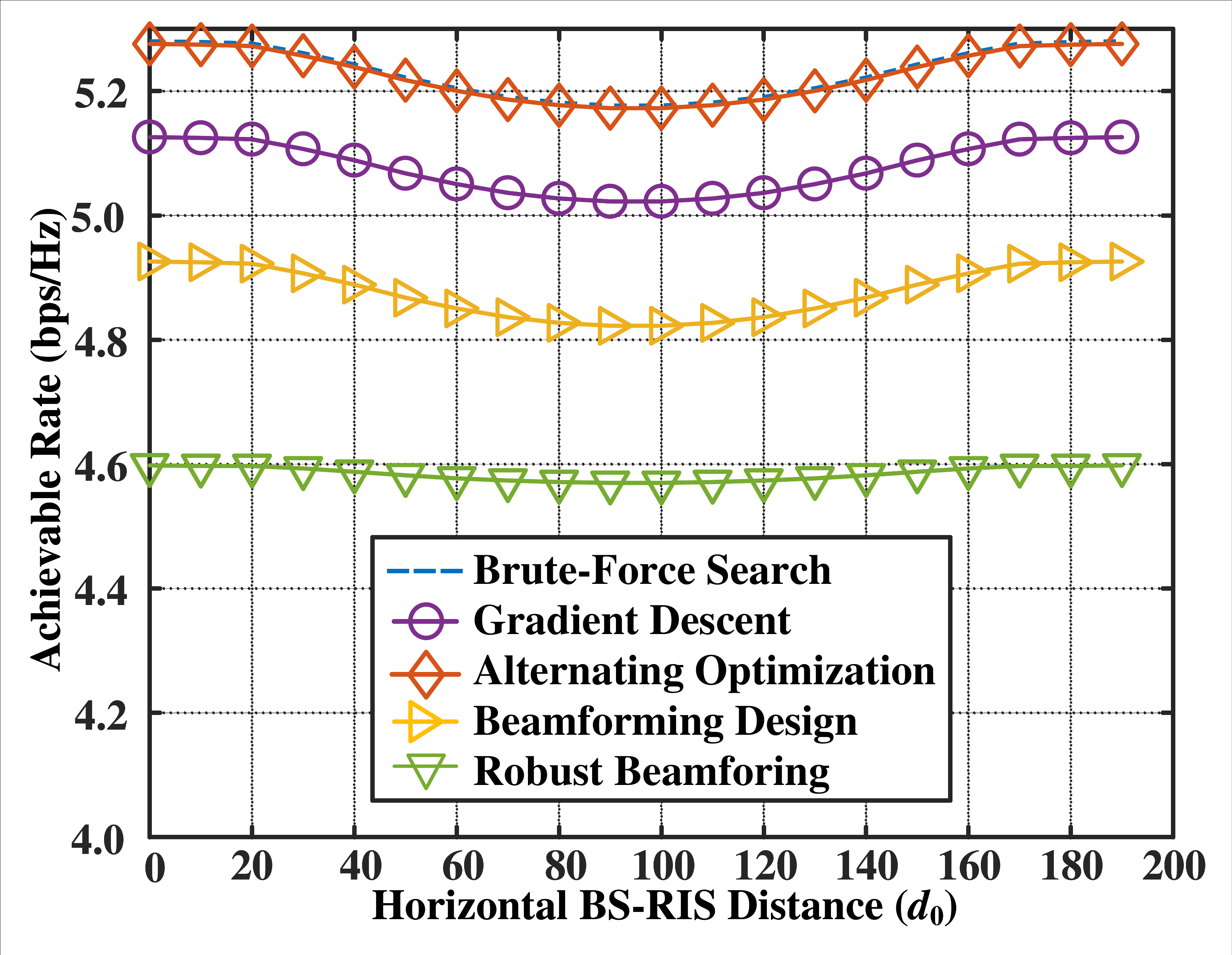}
\caption{Achievable rate versus the equal numbers of MUs ($K$) and antennas ($M$).}\label{fig_D}
\end{figure}

 Fig. \ref{fig_D} considers the effect of the path-loss in far-field communication.  The 3GPP propagation environment described in \cite{8741198}: 1) The large scale fading mode at distance $d$ is given as $\frac{10^{-3.75}}{d^{3.76}}$; 2) The BS and RIS are located at $\left(0,0\right)$ and $\left(d_0,50m\right)$, respectively, where $d_0$ is the horizontal BS-RIS distance, and the MUs are randomly and uniformly placed in the circular region with centre $\left(150m,0\right)$ and radius 50m. Moreover, we set $K=8$, $M=8$, $N=8$, $P_{\max}=10$ dBW, and the power of noises is set as $-80$ dBm.
From Fig.  \ref{fig_D}, it is concluded that the RIS near either the BS and MUs owns higher capacity than the RIS in the middle of BS and MUs. This is because the RIS near the BS can reflect a stronger signal from the BS, and the MU near the RIS is able to  receive the stronger reflected signal from the RIS.


\section{Conclusions}\label{section_conclusion}
 This paper considered a RIS assisted multicast communication. First, we formulated a capacity optimization problem and proposed two algorithms to obtain the locally optimal covariance matrix and phase shifts. Then, this paper considered a special case, where we only need to optimize the MU with the worst channel condition.  Lastly, we analyzed the order growth for the optimal capacity of RIS assisted  multicast transmission  when the numbers of reflecting elements, BS antennas, and MUs go to infinity.

\appendices

\section{Proof of Proposition \ref{Proposition_asymptotic_MN}}\label{proof_asymptotic_MN}
It is obvious that $C $ is smaller than the capacity of the channel from BS to any MUs, i.e.,
\begin{align}
C \leq\max_{\boldsymbol{\theta}}\log\left(1+P_{\max}\left \|  \mathbf{u}^{H}\mathbf{G}_k \mathbf{H}+\mathbf{t}_{k}^{H}\right \|^2 \right),
\end{align}
which is also an upper bound for $C $.  From (\ref{h}) and (\ref{H}), we obtain
\begin{align}
  \mathbf{u}^{H}\mathbf{G}_k \mathbf{H}+\mathbf{t}_{k}^{H}
=\left[\Psi_{1,k}, \Psi_{2,k},\dots,\Psi_{m,k}\right],
\end{align}
where $\Psi_{m,k}=\sum_{n=1}^{N}g_{k,n,m}+t_{k,m}$, with $g_{k,n,m}=\Phi_nh_{k,n} H_{n,m}$.
By the definitions in the first of section \ref{section_asymptotic}, we obtain $h_{k,n}\sim \mathcal{CN}\left(\sqrt{\frac{B}{B+1}}a_n,\frac{1}{B+1}\right)$, $H_{n,m}\sim \mathcal{CN}\left(\sqrt{\frac{B}{B+1}}a_na_m,\frac{1}{B+1}\right)$, $t_{k,m}\sim \mathcal{CN}\left(\sqrt{\frac{B}{B+1}}a_m,\frac{1}{B+1}\right)$, where $a_n=e^{j2\pi  \frac{d}{\lambda} (n-1)\sin^2 \omega\cos \vartheta\sin \vartheta}$.  Since $\left|\Phi_n\right|=1$, $\left|a_n\right|=1$ and $h_{k,n}$, $H_{n,m}$, $t_{k,m}$ are independent, it follows
\begin{align}
&\mathbb{E}\left\{ \left|\Psi_{m,k}\right|^2\right\}\nonumber\\
=&\left(N\frac{B}{B+1}+\sqrt{\frac{B}{B+1}}\right)^2+N\left(\left(\frac{B}{B+1}+\frac{1}{B+1}\right)^2-\left(\frac{B}{B+1}\right)^2\right)\nonumber\\
&+\frac{1}{B+1}\nonumber\\
=&A\left(N^2\right).\label{proof_hH_3}
\end{align}
 Then, based on the law of large numbers and (\ref{proof_hH_3}),  it follows
\begin{align}
\left \|  \mathbf{u}^{H}\mathbf{G}_k \mathbf{H}+\mathbf{t}_{k}^{H}\right \|^2&=\sum_{m=1}^{M} \left|\Psi_{m,k}\right|^2\label{proof__hH_law}\\
&\rightarrow  M \mathbb{E}\left\{ \left|\Psi_{m,k}\right|^2\right\} \nonumber\\
&=MA\left(N^2\right)\label{proof_MN_2},
\end{align}
a.s., as $M\rightarrow\infty$.
 Thus, $C$ is upper bounded by
\begin{align}
\log\left(1+P_{\max}MA\left(N^2\right)\right).\label{upper_bound_MN}
\end{align}
Next, from \cite{4036286}, it is obtained that $C $  is lower bounded in
\begin{align}
 \max_{\boldsymbol{\theta}}\log\left(1+\frac{P_{\max}}{K^2}\left \|  \mathbf{u}^{H}\mathbf{G}_k \mathbf{H}+\mathbf{t}_{k}^{H}\right \|^2 \right).
\end{align}
By the same proof for the upper bound of $C$, the lower bound of that can be rewritten as
 \begin{align}
&\max_{\boldsymbol{\theta}}\log\left(1+\frac{P_{\max}}{K^2}\left \| \mathbf{u}^{H}\mathbf{G}_k  \mathbf{H}+\mathbf{t}_{k}^{H}\right \|^2 \right)\nonumber\\
\rightarrow&\log\left(1+P_{\max}\frac{MA\left(N^2\right)}{K^2}\right),\label{lower_bound_MN}
\end{align}
a.s., as  $M\rightarrow\infty$.

From the upper and lower bounds in (\ref{upper_bound_MN}) and (\ref{lower_bound_MN}), we obtain that
\begin{itemize}
\item When $K$ and $N$ are fixed and $M$ goes to infinity, both the  upper and lower bounds of $C $ are  $\mathcal{O}\left(\log M\right)$, and it follows (\ref{Asymptotic_M});
\item When $N$ and $M$ go to infinity and $K$ is fixed, both the upper and lower bounds of $C $ are  $\mathcal{O}\left(\log N^2M\right)$, and it follows (\ref{Asymptotic_NM}).
\end{itemize}

Last, the proof of  (\ref{Asymptotic_N}) is similar to that of (\ref{Asymptotic_M}) and (\ref{Asymptotic_NM}), and thus we only show its difference. By the law of large number,  we can obtain that as $N\rightarrow\infty$,
 \begin{align}
\left|\Psi_{m,k}\right|^2=&\left(\sum_{i=1}^{N}g_{k,n,m}+t_{k,m}\right)\left(\sum_{n=1}^{N}\bar{g}_{k,i,m}+\bar{t}_{k,m}\right)\nonumber \\
           \rightarrow &\sum_{n=1}^{N}\sum_{i=1,i\neq n}^{N}\mathbb{E}\left\{g_{k,n,m}\bar{g}_{k,i,m}\right\}+ \sum_{n=1}^{N}\mathbb{E}\left\{\left|g_{k,n,m}\right|^2\right\}\nonumber\\
&+2\sum_{n=1}^{N}\mathbb{E}\left\{g_{k,n,m}\bar{t}_{k,m}\right\}+\mathbb{E}\left\{\left|t_{k,m}\right|^2\right\}\\
=&\left(N\frac{B}{B+1}+\sqrt{\frac{B}{B+1}}\right)^2+N\left(1-\left(\frac{B}{B+1}\right)^2\right)+\frac{1}{B+1}\nonumber\\
=&A\left(N^2\right),
\end{align}
 It follows $\left \|\mathbf{u}^{H}\mathbf{G}_k \mathbf{H}+\mathbf{t}_{k}^{H}\right \|^2=MA\left(N^2\right)$, as  $N\rightarrow\infty$. The other steps of this proof is same to the proof of (\ref{Asymptotic_M}) and (\ref{Asymptotic_NM}), and thus is omitted.

\section{Proof of Proposition \ref{Proposition_asymptotic_KMN}}\label{proof_asymptotic_K}

We first show the lower bound of $\mathbb{E}\left\{C \right\}$.
 From the proof of Propositions \ref{Proposition_asymptotic_MN}, we obtain that the mean of $\left\|\mathbf{u}^{H}\mathbf{G}_k \mathbf{H}+\mathbf{t}_{k}^{H}\right \|^2$ is $MA\left(N^2\right)$ and the variance of that is $MD\left(N^3\right)$, where
\begin{align}
D\left(N^3\right)=&4\left(\left(1-\left(\frac{B}{B+1}\right)^2\right)N+\frac{1}{B+1}\right)\left(\frac{NB}{B+1}+\sqrt{\frac{B}{B+1}}\right)^2\nonumber\\
&+2\left(\left(1-\left(\frac{B}{B+1}\right)^2\right)N+\frac{1}{B+1}\right)^2.
\end{align}
 It follows that the mean and variance of $\frac{\left \|\mathbf{u}^{H}\mathbf{G}_k \mathbf{H}\right \|^2}{M}$ are $A\left(N^2\right)$ and $\frac{D\left(N^3\right)}{M}$, respectively. Therefore, letting $l\leq\frac{1}{B+1}\left(N+2\right)/2+\frac{B}{B+1} \left(N+1\right)^2$, we derive
\begin{align}
&P\left\{\frac{\left\|\mathbf{u}^{H}\mathbf{G}_k \mathbf{H}+\mathbf{t}_{k}^{H}\right \|^2}{M}\leq l\right\}\nonumber\\
=&P\left\{\frac{\left\|\mathbf{u}^{H}\mathbf{G}_k \mathbf{H}+\mathbf{t}_{k}^{H}\right \|^2}{M}-A\left(N^2\right)\leq\left(l-A\left(N^2\right)\right)\right\}\nonumber\\
\leq & P\left\{\left|\frac{\left\|\mathbf{u}^{H}\mathbf{G}_k \mathbf{H}+\mathbf{t}_{k}^{H}\right \|^2}{M}-A\left(N^2\right)\right|\geq\left(A\left(N^2\right)-l\right)\right\}\nonumber\\
\leq &\frac{D\left(N^3\right)}{M\left(A\left(N^2\right)-l\right)^2},\label{cheby_1}
\end{align}
where (\ref{cheby_1}) is due to Chebychev inequality. By the extreme value theory \cite{fisz2018probability}, we obtain
\begin{align}
&P\left\{\min_{k=1,\cdots,K}\frac{\left\|\mathbf{u}^{H}\mathbf{G}_k \mathbf{H}+\mathbf{t}_{k}^{H}\right \|^2}{M}\geq l\right\}\nonumber\\
=& \left(1-P\left\{\frac{\left\|\mathbf{u}^{H}\mathbf{G}_k \mathbf{H}+\mathbf{t}_{k}^{H}\right \|^2}{M}\leq l\right\}\right)^K.\label{extreme_theory}
\end{align}
Then, combing (\ref{cheby_1}) and (\ref{extreme_theory}), it is obtained that
\begin{align}
&P\left\{\min_{k=1,\cdots,K}\frac{\left\|\mathbf{u}^{H}\mathbf{G}_k \mathbf{H}+\mathbf{t}_{k}^{H}\right \|^2}{M}\geq l\right\}\nonumber\\
=& \left(1-\frac{D\left(N^3\right)}{M\left(A\left(N^2\right)-l\right)^2}\right)^K\nonumber\\
\rightarrow& \exp\left(-\frac{KD\left(N^3\right)}{M\left(A\left(N^2\right)-l\right)^2}\right)\label{lower_bound_KM_previous},
\end{align}
as $K\rightarrow \infty$. It is obvious that the lower bound of $\mathbb{E}\left\{C \right\}$ is given as
\begin{align}
\mathbb{E}\left\{C \right\}&\geq \mathbb{E}\log\left(1+P_{\max}\min_{k=1,\cdots,K}\frac{\left\|\mathbf{u}^{H}\mathbf{G}_k \mathbf{H}+\mathbf{t}_{k}^{H}\right \|^2}{M}\right)\nonumber\\
&\geq P\left\{\min_{k=1,\cdots,K}\frac{\left\|\mathbf{u}^{H}\mathbf{G}_k \mathbf{H}+\mathbf{t}_{k}^{H}\right \|^2}{M}\geq l\right\}\log\left(1+l P_{\max}\right)\nonumber\\
& \rightarrow  \exp\left(-\frac{KD\left(N^3\right)}{M\left(A\left(N^2\right)-l\right)^2}\right) \log\left(1+l P_{\max}\right)\label{lower_bound_KM},
 \end{align}
as $K\rightarrow \infty$, where (\ref{lower_bound_KM}) is due to (\ref{lower_bound_KM_previous}). From (\ref{lower_bound_KM}), it is easy to see that since $l<A\left(N^2\right)$ and $\frac{K}{M}\leq \infty$  is a fixed ratio, the lower bounds of $\mathbb{E}\left\{C \right\}$ are both $\mathcal{O}\left(1\right)$ for the case of fixed $N$ and $N\rightarrow \infty$ cases, respectively.

Next, we show the upper bound of $\mathbb{E}\left\{C \right\}$.  For the minimum received  SNR, we derive
\begin{align}
&\max_{\mathcal{Q}}\min_{k=1,\cdots,K} \left(\mathbf{u}^{H}\mathbf{G}_k \mathbf{H}+ \mathbf{t}_{k}^{H}\right)\mathbf{Q}\left(\mathbf{H}^{H}\mathbf{G}^{H}_k \mathbf{u}+\mathbf{t}_{k}\right)\nonumber\\
\leq &\frac{1}{K}\max_{\mathcal{Q}}\sum_{k=1}^{K}\left(\mathbf{u}^{H}\mathbf{G}_k \mathbf{H}+ \mathbf{t}_{k}^{H}\right)\mathbf{Q}\left(\mathbf{H}^{H}\mathbf{G}^{H}_k \mathbf{u}+\mathbf{t}_{k}\right)\label{uppwer_bound_K_3}\\
=&\frac{1}{K}\max_{\mathcal{Q}}\mathrm{Tr}\left(\mathbf{Q}\mathbf{\Omega}\mathbf{\Omega}^H \right)\label{uppwer_bound_K_1}\\
=&\frac{P_{\max}}{K}\lambda_{\max}\left(\mathbf{\Omega}\mathbf{\Omega}^H \right)\label{uppwer_bound_K_2},
 \end{align}
where $\mathbf{\Omega}=\left[\mathbf{H}^{H}\mathbf{G}^{H}_1 \mathbf{u}+\mathbf{t}_{1},\cdots,\mathbf{H}^{H}\mathbf{G}^{H}_k \mathbf{u}+\mathbf{t}_{K}\right]$; (\ref{uppwer_bound_K_3}) is due to the fact that the minimum received SNR is upper bounded by the average received SNR;  (\ref{uppwer_bound_K_2}) is due to the fact that the maximization (\ref{uppwer_bound_K_1}) is equivalent to the maximal eigenvalue of the matrix $\lambda_{\max}\left(\mathbf{\Omega}\mathbf{\Omega}^H\right)$\cite{IEEEhowto:S.Boyd}. From the eigenvalue of random matrix theory \cite{JackW1985}  ($\Psi_{m,k}$ is elements of matrix $\mathbf{\Omega}$), we obtain
\begin{align}
\frac{P_{\max}}{K}\lambda_{\max}\left(\mathbf{\Omega}\mathbf{\Omega}^H \right)\rightarrow P_{\max}A\left(N^2\right)\left(1+\sqrt{\frac{M}{K}}\right)^2,
 \end{align}
a.s.,  as $K\rightarrow \infty$, $M\rightarrow \infty$ and $\frac{K}{M}<\infty$, which implies
\begin{align}
C \leq \log\left(1+ P_{\max}A\left(N^2\right)\left(1+\sqrt{\frac{M}{K}}\right)^2\right).\label{upper_bound_KM}
 \end{align}
It follows $\mathbb{E}\left\{C \right\}\leq \mathcal{O}\left(1\right)$ for the fixed $N$, and $\mathbb{E}\left\{C \right\}\leq \mathcal{O}\left(\log(N^2)\right)$ for the $N\rightarrow \infty$.

\bibliographystyle{IEEEtran}
\bibliography{Full_mul}

\end{document}